\documentclass[preprint,onecolumn,nofootinbib,12pt]{revtex4}
\usepackage{graphicx}
\pdfoutput=1
\usepackage{amsmath}
\usepackage{amsfonts}
\usepackage{amssymb}
\usepackage{color}
\usepackage{dcolumn}
\usepackage{hyperref}
\usepackage{bm}
\usepackage{ulem}
\providecommand{\U}[1]{\protect\rule{.1in}{.1in}}

\newcommand{\f}{\begin{equation}}
\newcommand{\ff}{\end{equation}}
\newcommand{\fa}{\begin{eqnarray}}
\newcommand{\ffa}{\end{eqnarray}}

\begin{document}
\title{Holographic Butterfly Effect and Diffusion in Quantum Critical Region}
\author{Yi Ling $^{1,2}$}
\email{lingy@ihep.ac.cn}
\author{Zhuo-Yu Xian $^{1,2}$}
\email{xianzy@ihep.ac.cn}
\affiliation{$^1$ Institute of High Energy Physics, Chinese Academy of Sciences, Beijing 100049, China\ \\
$^2$ School of Physics, University of Chinese Academy of Sciences, Beijing 100049, China}

\begin{abstract}
We investigate the butterfly effect and charge diffusion near the quantum phase transition in holographic approach. We argue that their criticality is controlled by the holographic scaling geometry with deformations induced by a relevant operator at finite temperature. Specifically, in the quantum critical region controlled by a single fixed point, the butterfly velocity decreases when deviating from the critical point. While, in the
non-critical region, the behavior of the butterfly velocity depends on the specific phase at low temperature. Moreover, in the holographic Berezinskii-Kosterlitz-Thouless transition, the universal behavior of the butterfly velocity is absent. Finally, the tendency of our holographic results matches with the numerical results of Bose-Hubbard model. A comparison between our result and that in the $O(N)$ nonlinear sigma model is also given.

\end{abstract}
\maketitle

\section{Introduction}
\subsection{Background}
Quantum chaos is a fascinating phenomenon and plays a key role in
understanding thermalization in many-body system. Two general
processes related to chaos are relaxation and scrambling. As a
characteristic quantity describing relaxation, the relaxation time
$\tau_\text{relax}$ can be calculated by the local decay of
time-order two point function \cite{Forster:1975hyd}. When
temperature $T$ becomes the dominant scale in a system, the
relaxation time scales with the temperature as
$\tau_\text{relax}\sim T^{-1}$, where $\hbar=k_B=1$. Usually,
relaxation is followed by scrambling, during which the
scrambling time $t_*$ measures the time for a system to lose the
memory of its initial state \cite{Hosur:2015ylk,Sekino:2008he}.
Black hole has the fastest scramble process and performs chaos in the decrease of the mutual information after a perturbation \cite{Sekino:2008he,Shenker:2013pqa,Sircar:2016old,Cai:2017ihd}.
For a gauge theory with rank $N$, the scrambling time
behaves as $t_*\sim \tau_L \log N^2$
\cite{Maldacena:2015waa,Sekino:2008he}, where $\tau_L$ is the
Lyapunov time defined by the reciprocal of Lyapunov exponent
$\lambda_L$ as $\tau_L\equiv1/\lambda_L$. While Lyapunov exponent
$\lambda_L$ can be extracted from the square of the commutator
\cite{Kitaev:2014hch,Maldacena:2015waa,Shenker:2013pqa,Roberts:2014isa,Roberts:2016wdl,Roberts:2014ifa,Shenker:2013yza,Stanford:2015owe,Maldacena:2016hyu,Polchinski:2016xgd}
\begin{equation}\label{Commutator}
  \tilde C(t,x)=\langle[W(x,t),V(0)]^\dag[W(x,t),V(0)]\rangle_\beta \sim A_C
  e^{\lambda_L(t-t_*-|x|/v_B)},
\end{equation}
where $W(x,t)=e^{iHt}W(x)e^{-iHt}$ and $W(x)$ and $V(0)$ are local
operators at $x$ and $0$. $\langle\cdots\rangle_\beta\equiv
Z^{-1}Tr\{e^{-\beta H}\cdots\}$ denotes the ensemble average at
temperature $T=\beta^{-1}$ and $A_C$ is a normalized factor. For a
gauge theory with rank $N$, one has $A_C\sim N^{-2}$. The
commutator $\tilde C(t,x)$ become significant at scrambling time
$t_*$.

The Lyapunov exponent $\lambda_L$ \cite{Kitaev:2014hch}
characterizes how chaos grow for early time. Similar to the
Kovtun-Son-Starinets (KSS) bound for $\eta/s$
\cite{Kovtun:2004de}, Maldacena {\it et. al.}
\cite{Maldacena:2015waa} conjectured a universal bound on chaos,
\begin{equation}\label{ChaosBound}
\lambda_L\leq2\pi T,
\end{equation}
which is saturated in Einstein gravity and Sachdev-Ye-Kitaev (SYK)
model \cite{Kitaev:2014hch}. It is further conjectured that a
large-$N$ system will have an Einstein gravity dual in the near
horizon region if the bound (\ref{ChaosBound}) is saturated
\cite{Kitaev:2014hch,Maldacena:2015waa}. Unlike the KSS bound, the
bound for $\lambda_L$ is unchanged even in gravity theories with
higher derivative corrections
\cite{Kitaev:2014hch,Maldacena:2015waa}.

Butterfly velocity $v_B$ characterizes how chaos spreads in space
\cite{Shenker:2013pqa}. One can define a `butterfly' cone,
$t-|x|/v_B = t_*$, inside the light cone \cite{Roberts:2014isa}.
For unitary operators $W(x)$ and $V(0)$, the normalized
commutator $\tilde C(t,x)/(\langle W W\rangle_\beta\langle V
V\rangle_\beta)$ is nearly zero outside the butterfly cone, which
means that the part of system is not affected by the perturbation
of $V(0,0)$ \cite{Roberts:2016wdl}. Later, when crossing the
butterfly cone it exponentially increases. At the final
stage, it saturates the value $2$ inside the butterfly cone and
the exponential behavior in (\ref{Commutator}) breaks down
\cite{Roberts:2014isa}.

Out-of-time-order correlation (OTOC) function plays a similar role
in the study of chaos, which is defined as
\cite{Kitaev:2014hch,Maldacena:2015waa,Shenker:2013pqa,Roberts:2014isa,Roberts:2016wdl,Roberts:2014ifa,Shenker:2013yza,Stanford:2015owe,Maldacena:2016hyu,Polchinski:2016xgd}
\begin{equation}\label{OTOC}
\tilde F(t,x)=\langle
W^\dag(t,x)V^\dag(0,0)W(t,x)V(0,0)\rangle_\beta \sim \alpha_0 -
\alpha_1 e^{\lambda_L(t-|x|/v_B)}.
\end{equation}
It is linked to (\ref{Commutator}) by $\tilde
C(t,x)=2(1-\text{Re}[\tilde F(t,x)])$ when $W(x)$ and $V(0)$ are
unitary operators \cite{Swingle:2016var,Roberts:2014isa}.

Usually, it is rather complicated to calculate OTOC in a
many-body system \cite{Maldacena:2016hyu,Polchinski:2016xgd}.
Thanks to the gauge/gravity duality, recent progress
indicates that the holographic nature of gravity may shed light on
quantum butterfly effect which can be viewed as a dual of
shockwave solutions in an asymptotically AdS black hole background
\cite{Kitaev:2014hch,Shenker:2013pqa,Shenker:2013yza,Roberts:2014isa}.
This directly stimulates us to further investigate the butterfly
effect in holographic approach in this paper, with a focus on its
behavior close to the quantum critical point.

On the other hand, motivated by the charge diffusion bound on
incoherent metal \cite{Hartnoll:2014lpa}, Blake
\cite{Blake:2016wvh,Blake:2016sud} recently proposed that $v_B$ may work as the
characteristic velocity bounding diffusion constant $D$ in incoherent transport \footnote{We thank Wei-Jia Li for clarifying the applicability of such bound and drawing our attention to the momentum transport.},
\begin{equation}\label{OriginalDiffusionBound}
D\gtrsim v_B^2/T.
\end{equation}
Here, the symbol `$\gtrsim$' means greater up to a
constant. The most concerned diffusion constants contain the
charge diffusion constant $D_c$, the energy diffusion constant
$D_e$ and the momentum diffusion constant $D_p$, when their
diffusive quantities are conserved. Blake's conjuncture has been
tested in many holographic models
\cite{Blake:2016wvh,Lucas:2016yfl,Patel:2016wdy,Kim:2017dgz,Blake:2016jnn,Baggioli:2016pia,Baggioli:2017ojd,Hartman:2017hhp,Blake:2017qgd,Blake:2016sud}
and condensed matter models
\cite{Gu:2017ohj,Davison:2016ngz,Gu:2016oyy}. The bound for $D_c$
is found to be violated in
\cite{Lucas:2016yfl,Baggioli:2016pia,Davison:2016ngz}. A possible
explanation for the violation is that chaos should be linked to
the loss of quantum coherence and energy fluctuations, rather than
the transportation of conserved electric charges
\cite{Patel:2016wdy,Davison:2016ngz,Gu:2017ohj}. Recently, a
stronger bound for energy diffusion constant,
\begin{equation}\label{StrongerDiffusionBound}
D_e\gtrsim v_B^2\tau_L,
\end{equation}
is studied in \cite{Gu:2017ohj,Bohrdt:2016vhv}. When a quantum
field theory has a holographic dual
\cite{Blake:2016wvh,Lucas:2016yfl,Blake:2016jnn,Kitaev:2014hch,Gu:2017ohj,Blake:2016sud},
the bound for $\lambda_L$ in (\ref{ChaosBound}) is saturated and
then the original bound (\ref{OriginalDiffusionBound}) and
stronger bound (\ref{StrongerDiffusionBound}) are equivalent.
However, such stronger bound (\ref{StrongerDiffusionBound}) is
violated in inhomogeneous SYK chains \cite{Gu:2017ohj}, which
raises a puzzle on the relation between transport and quantum
chaos in strange metals.

\subsection{Butterfly effect near the quantum critical point}
Inspired by recent progress in holography, OTOC
(\ref{OTOC}) has been studied near quantum phase transition
(QPT) in many-body systems
\cite{Shen:2016htm,Patel:2016wdy,Bohrdt:2016vhv,Chowdhury:2017jzb}.
In \cite{Shen:2016htm}, Shen {\it et. al.} found that both
$\lambda_L$ and $v_B$ reach a maximum near the critical
point $g=g_c$ at finite temperature in $(1+1)$ dimensional
Bose-Hubbard model (BHM), XXZ model and transverse Ising model.
They also conjectured that $\lambda_L$ would display
a maximum around the quantum critical point (QCP), which is
equivalent to the minimization of Lyapunov time $\tau_L$.

Critical phenomenon is a very nice area for observing the
universality of a system because the microscopic details become
irrelevant near the critical point. It is quite nature to expect
that the bounds and extremal behaviors mentioned above for
butterfly effect and diffusion would also exhibit some universal
feature during phase transitions. To better understand this, we
intend to briefly review the basic structure in quantum critical
phenomenon, which takes place in continuous QPT.
A general QPT can be accessed by tuning some coupling
constant $g$ crossing a critical point $g_c$ at zero
temperature $T=0$ \cite{Sachdev:1999QPT}. If such QPT is
continuous, the point $g_c$ is called the QCP, at
which the correlation length $\xi$ diverges. In particular,
if the QPT is the second order, then from the viewpoint of
renormalization group (RG), the QCP corresponds to an unstable
fixed point, which enjoys the property of scaling invariance. Here
for scale transformations, we remark that time and space may have
different scaling dimensions
\begin{equation}
[t]=-z, \, [x]=-1,
\end{equation}
where $z$ is the dynamical critical exponent.

There are two important scales near the QCP, namely, the
temperature $T$ and the distance $g-g_c$ away from QCP, whose
scaling dimensions are separately given as
\begin{equation}
  [T]=z, \, [g-g_c]=1/\nu,
\end{equation}
where $\nu$ is another critical exponent. Both of the scaling
dimensions should be positive, ensuring that their
deformations to the QCP are relevant\footnote{So far we
have only considered the QCP with hyperscaling symmetry,
i. e. the hyperscaling violating exponent $\theta=0$.
So a relevant thermal deformation requires $z>0$. Actually,
even hyperscaling is violated, the region of $z<0$ is found to be
`pathological' from the perspective of the consistent
dimensional reduction and entanglement entropy
\cite{Dong:2012se,Gouteraux:2011qh,Gouteraux:2011ce}.}. From the perspective of quantum
field theory (QFT), the scale $g-g_c$ can be
introduced by deforming the fixed point theory with a
relevant operator
\cite{Hartnoll:2016apf,Sachdev:1999QPT,Lucas:2017dqa,Hartnoll:2009sz,Sachdev:2010ch}
\begin{equation}\label{ActionDeform}
 S_\text{QFT} = S_\text{fixed point} + \kappa \int dx^{d+1} W(\cal
  O),
\end{equation}
where $W(\cal O)$ is a function of operator $\cal O$ and the
source $\kappa$ is identified with $g-g_c$, namely $\kappa\sim
g-g_c$. Hence, $[\kappa]=1/\nu$. In addition, for a QPT there
exists an UV scale $\Lambda_\text{UV}$, which is close to
the energy of microscopic interaction in a many-body
system. When $T\gg\Lambda_\text{UV}$, it is called the region of
lattice high temperature \cite{Sachdev:1999QPT}. Quantum critical
phenomenon emerges when $T\ll \Lambda_\text{UV}$. The competition
between two scales $T$ and $\kappa$ divides the phase diagram into
the quantum critical region and the non-quantum-critical region,
as illustrated in the left plot of Figure \ref{FigQPT}.

In the quantum critical region, $T\gg|\kappa|^{z\nu}$, temperature
$T$ is the dominant scale. In general, under external
perturbations a system will lose local quantum phase coherence and
such a process can be characterized by the phase coherent time
$\tau_\varphi$, which can technically be evaluated by the
exponential decay of local commutators, which is close to the measurement of relaxation time
$\tau_\text{relax}$. A general feature about the phase coherent
time $\tau_\varphi$ is
\begin{equation}\label{coherencetime}
  \tau_\varphi\gtrsim T^{-1},
\end{equation}
which becomes saturated in quantum critical region
\cite{Sachdev:1999QPT,Hartnoll:2016apf}.
Therefore, in this region both of quantum and thermal
fluctuations are important, which usually leads to a
non-quasiparticle description of dynamics at finite temperature.

In the non-quantum-critical region where
$T\ll|\kappa|^{z\nu}$, the specific low temperature phase is
controlled by the corresponding IR fixed point of theory
(\ref{ActionDeform}). For a gapped phase there are quasi-particles
with energy gap $\Delta_E\sim|\kappa|^{z\nu}\gg T$, leading to
sparse excitations and a long phase coherence time
$\tau_\varphi\sim e^{\Delta_E/T}$.

The similarity between the bound for $\tau_L$ in
(\ref{ChaosBound}) and the bound for $\tau_\varphi$ in
(\ref{coherencetime}) has been suggested in
\cite{Hartnoll:2016apf}. The conjecture about the minimization of
Lyapunov time $\tau_L$ in \cite{Shen:2016htm} is also
reminiscent of (\ref{coherencetime}). Therefore, in this
sense it is very desirable to check the scaling behaviors of
$\tau_L$ in the phase diagram of QPT. Recently, the scaling of
$\tau_L$ is calculated in the $O(N)$ nonlinear sigma model at
large $N$ in \cite{Chowdhury:2017jzb}, where it is found that
$\tau_L\sim T^{-1}$ in the quantum critical region, $\tau_L\sim
T^{-3}$ in the symmetry-broken region and $\tau_L\sim
e^{2\Delta_E/T}$ in the symmetry-unbroken region. However, the
value of $\tau_L$ obtained from the side of classical
gravity always saturates the bound in (\ref{ChaosBound}). So it is
difficult to test the conjecture on the minimization of $\tau_L$
in the quantum critical region in holographic approach.

As seen from above, the dominant scale $T^{-1}$ bounds other
 dynamical time scales in the quantum critical
region\footnote{It is so called `Planck time' $t_\text{pl}\equiv
T^{-1}$ in \cite{Zaanen:2004sth}.}. However, when another
important scale $\kappa$ is involved, one may expect deviations of
those time scales from $T^{-1}$. From (\ref{coherencetime}),
it is understood that $\tau_\varphi$ has to increase and
deviate from $T^{-1}$ when leaving the quantum critical region at
fixed temperature $T$. Nevertheless, (\ref{coherencetime})
is just an approximate description, and does not guarantee
that $\tau_\varphi$ must reach a minimum at $\kappa=0$. We
will discuss this in subsection \ref{SubsectionONModel}.

After having discussed the time scales in QPT, we turn
back to the butterfly effect. What kind of behavior
should we expect for $v_B$ near QCP? To answer this
question, let us firstly estimate the characteristic velocity
$v_{qp}$ of quasi-particles in some weakly coupled
many-body system. For example, we assume a relativistic dispersion
relation $\epsilon_{k}^2=c^2{k}^2+m_{qp}^2$, where $c$ is the
speed of light and $m_{qp}$ is the effective mass of
quasi-particles. At high temperature, $T\gg m_{qp}$, from
the estimation $T\approx\epsilon_k$, we have
\begin{equation}\label{vqp}
  v_{qp} = \frac{\partial \epsilon_k}{\partial k} \approx c\sqrt{1-\left(\frac{m_{qp}}{T}\right)^2} \approx c\left(1-\frac12\left(\frac{m_{qp}}{T}\right)^2\right) + \cdots,
\end{equation}
where $T\gg m_{qp}$ has been applied in the last
approximate equality. One simple but direct interpretation
on (\ref{vqp}) is that the effective mass hinders the spread of
quasi-particles. If weak chaos can develop from the weak
interaction among quasi-particles, we expect a decrease of $v_B$ similar to (\ref{vqp}).

Now let us discuss $v_B$ in the quantum critical region. As we
have mentioned before, quasi-particle usually is not well defined,
let alone its velocity $v_{qp}$. Nevertheless, based
on the spirit in \cite{Blake:2016wvh}, $v_B$ should be able to
stand for a characteristic velocity even in the quantum critical
region, no matter the QCP is relativistic or not. Unlike
(\ref{vqp}), without quasi-particle scenario, the
calculation of $v_B$ will be complicated and the dependence
of $v_B$ on two scales $T$ and $\Delta_E$ is not clear on
field theory side. However, we still expect some
similarity between the weakly coupled
system and the near critical system. Specifically, $m_{qp}$ should
correspond to $\Delta_E$, as $m_{qp}$ is just the energy gap of
quasi-particle-like excitations. So the relation $T\gg
m_{qp}$ would correspond to $T\gg\Delta_E$, i. e. the
condition of the quantum critical region. We will see that such
naive correspondences are consistent with the result of
$v_B$ (\ref{vBqcr}) from holography.

For the estimation of $v_B$ in non-quantum-critical region, the
picture of quasi-particle is useful on field theory side, see
\cite{Chowdhury:2017jzb}. While, on gravity side, the strategy is
different. In \cite{Ling:2016ibq}, $v_B$ displays distinct
scaling behaviors in different phases. Thus a discontinuity
of $v_B$ appears close to the critical point at rather low
temperature, which leads to a peak of $\partial v_B/\partial g$ as
well. Since such phenomenon is controlled by two fixed
points separately corresponding to low temperature phases,
while $v_B$ in the quantum critical region is controlled by the
dynamics of QCP, its behavior in these different regions has
no direct connections.

\subsection{Scaling formula for the butterfly velocity and diffusion
constant}

Finally, it is interesting to understand the charge
diffusion bound (\ref{OriginalDiffusionBound}) in the quantum
critical region. Based on dimensional analysis, our direct
expectation is following. When $\kappa$ vanishes, $v_B$ can be
written as $v_B^2\sim T^{2-\frac2z}$ according to its scaling
dimension $[v_B]=[x]-[t]=-1+z$ \footnote{Dimensional analysis does
not work in the generalized SYK model and the holographic theories
with $AdS_2\times R^d$ near horizon geometry, since spaces are
decoupled from the scaling symmetry. By further considering
the spatial irrelevant modes, it is found that $v_B\sim \sqrt{T}$
\cite{Gu:2016oyy,Blake:2016jnn}.}. When $\kappa$ is turned on, we
expect a scaling formula
\begin{equation}\label{VBscale}
  v_B^2=T^{2-\frac2z}\Phi\left(\frac{\kappa}{T^\frac1{z\nu}}\right),
\end{equation}
where $\Phi(x)$ is a function that should be determined by the
details of theory, and is expected to exhibit some
universal behavior when $x$ is small. For later convenience, we
always write out the expression for $v_B^2$ rather than
$v_B$ itself. Similar consideration can be applied to the charge
diffusion bound (\ref{OriginalDiffusionBound}), giving rise to a
dimensionless ``diffusion ratio''
\begin{equation}\label{Diffusionscale}
  \frac{D_c\lambda_L}{v_B^2}=\Psi\left(\frac{\kappa}{T^\frac1{z\nu}}\right),
\end{equation}
where $\Psi(x)$ is a function to be determined by the
theory as well.

In this paper, we are going to derive the specific forms of
(\ref{VBscale}) and (\ref{Diffusionscale}) near QCP for a
class of holographic models with classical gravity \cite{Hartnoll:2009sz,Sachdev:2010ch,Hartnoll:2016apf,McGreevy:2009xe}. So the dual
system under consideration is described by a large $N$ gauge
theory with strong couplings. Such kind of field theory
 at fixed point $S_\text{fixed point}$ in
(\ref{ActionDeform}) with dynamical critical exponent $z$ is dual
to the Lifshitz spacetime
\cite{Kachru:2008yh,Taylor:2008tg,Taylor:2015glc}. For large $N$
limit and without string corrections, $v_B$ and $D_c$ can be
calculated over a classical bulk geometry with the use of the
method developed in
\cite{Blake:2016jnn,Blake:2016wvh,Blake:2016sud}. Especially, we
focus on the quantum critical region, which is dominantly
controlled by the dynamics of the
QCP. The gravity dual at finite temperature
can be a black hole or a thermal gas \cite{Sachdev:2008ba}.

\begin{figure}
  \centering
  \includegraphics[height=100pt]{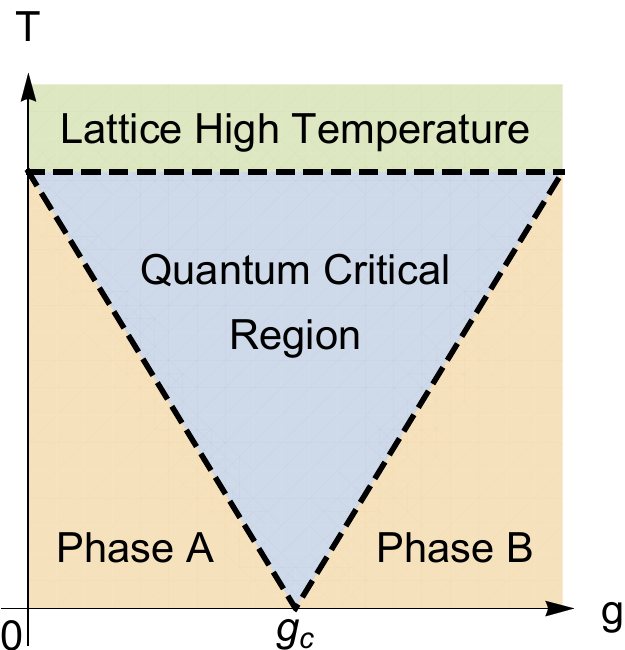}\quad
  \includegraphics[height=100pt]{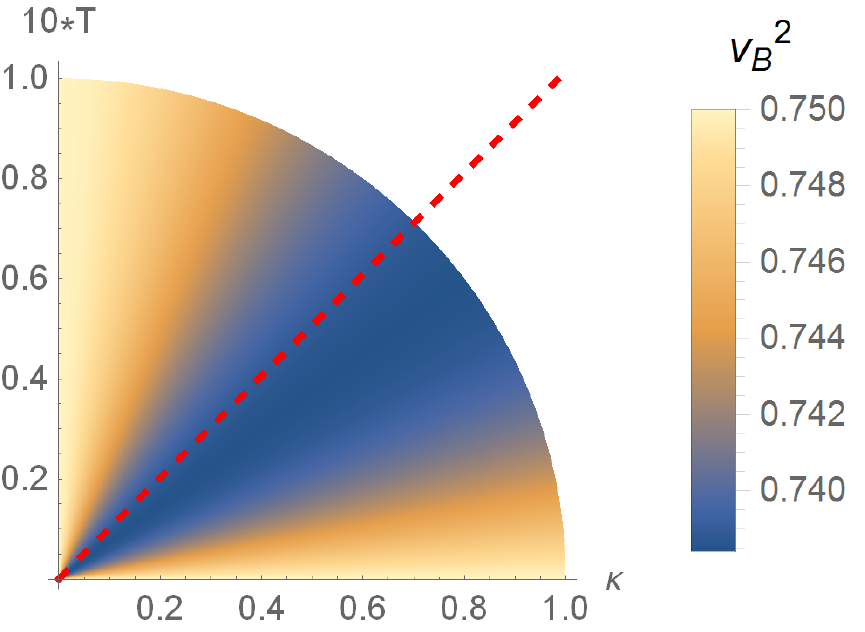}\\
  \caption{Left: The schematic phase diagram of quantum phase transition. Right: Density plot of butterfly velocity $v_B$ in the phase diagram of AdS-AdS domain wall. The red dashed line mark the place where scalar field at the horizon reaches half of its value at IR fixed point. Such line indicate the vague boundary between quantum critical region and non-quantum-critical-region, namely $T$ and $|\kappa|^{z\nu}$ are comparable ($z\nu$ in these two plots is equal to 1). One can compare these two plots by identifying $\kappa\sim g-g_c.$}\label{FigQPT}
\end{figure}

Before going into the details of the holographic construction, we
demonstrate an intuitive picture for the butterfly velocity
$v_B^2$ over the phase diagram with QPT obtained holographically.
As an illustration of the scaling formula in (\ref{VBscale}), we
numerically calculate $v_B^2$ over an AdS-AdS domain wall
background which is given in Appendix \ref{AppendixAdSNum}.
Its value over the phase diagram is shown in the right plot of
Figure \ref{FigQPT}. One can compare it with the schematic phase
diagram of QPT in the left plot by identifying $\kappa\sim g-g_c$.
The AdS-AdS domain wall is linked by a scalar $\phi$ which is dual
to the operator $\cal O_\phi$ in (\ref{ActionDeform}). The quantum
critical region, $T\gg|\kappa|^{z\nu}$, is dual to the UV AdS
black hole deformed by scalar field $\phi$; while the
non-quantum-critical-region, $T\ll|\kappa|^{z\nu}$, is dual to the
IR AdS black hole deformed by scalar field $\phi$. The nontrivial
IR AdS fixed point can be understood as an example of a gapless
phase. However, more common phases in QPT are gapped phases
flowing to trivial fixed points \cite{Sachdev:1999QPT}.

The isolation between the fixed points makes our aim clear. We
will mainly focus on the quantum critical region and do not need
to care about the IR fixed point to which the deformation $W(\cal
O_\phi)$ will drive the system.

We organize this paper as follows. In section
\ref{SectionAdS}, we calculate (\ref{VBscale}) and (\ref{Diffusionscale}) in an AdS
space with scalar field $\phi$ deformation,
which is dual to a conformal fixed point
with scalar operator deformation $W(\cal O_\phi)$. In section \ref{SectionLif}, we numerically calculate (\ref{VBscale}) in a Lifshitz fixed point with scalar deformation.  When $W(\cal
O_\phi)$ is a single trace deformation, we derive $v_B$ in quantum critical region as
\begin{equation}\label{vBqcr}
  v_B^2=T^{2-\frac2z}\Phi(0)\left(1-\gamma\frac{\kappa^2}{T^\frac2{z\nu}} + \cdots \right),\quad T\gg
  |\kappa|^{z\nu},
\end{equation}
where $\gamma$ is a non-negative constant independent of $T$
and $\kappa$. Therefore, this formula exhibits an universal
behavior that $v_B$ is always decreasing when the system is deformed away from $\kappa=0$ at fixed $T$.
In other words, $v_B$ reaches its local peak near the QCP. Surprisingly, the holographic result (\ref{vBqcr}) has the similar form as (\ref{vqp}) at $z=1$ if we recognize $\kappa^{\nu}\sim\Delta_E$ and replace $m_{qp}$ by $\Delta_E$ in (\ref{vqp}). We also calculate the charge
diffusion constant $D_c$ and the diffusion ratio
(\ref{Diffusionscale}) in the AdS case. The result for $d>1$ is
\begin{equation}\label{DcovB2tauqcr}
  \frac{D_c\lambda_L}{v_B^2}=\frac{d}{d-1}\left(1+\eta\frac{\kappa^2}{T^\frac2{\nu}}+\cdots\right) ,\quad T\gg
  |\kappa|^{\nu},
\end{equation}
where $\eta$ is a non-negative constant independent of $T$
and $\kappa$ as well. It shows that $D_c$ is bounded from below
by $\frac{d}{d-1}\frac{v_B^2}{\lambda_L}$ in the quantum critical
region and the ratio will increase when the system goes away
from $\kappa=0$. A similar increase is found in the ratio $D_p\lambda_L/v_B^2$ as well. In section \ref{SectionOther},
we turn to discuss $v_B$ in some other low temperature phases in
holography. Moreover, a simple holographic
Berezinskii-Kosterlitz-Thouless (BKT) phase transition will be
studied, as a comparison to the second order QPT in the main
text. In section \ref{SectionCMT}, we will compare our holographic
results with those from (1+1) Bose-Hubbard model in
\cite{Shen:2016htm,Bohrdt:2016vhv} and $(2+1)$ $O(N)$ non-linear
sigma model at large $N$ in \cite{Bohrdt:2016vhv}. Some
similarities and differences are found.

\section{AdS-Schwarzschild black hole with scalar deformation}\label{SectionAdS}
In this section we will investigate how the butterfly
velocity $v_B$ (\ref{VBscale}) and the charge diffusion ratio $D_c\lambda_L/v_B^2$ (\ref{Diffusionscale}) change under the scalar deformation
$W(\cal O_\phi)$. As the starting point, we consider a classical
Einstein gravity theory with the solution of $AdS_{d+2}$
spacetime, which is dual to a large $N$ and strongly coupled
theory at conformal fixed point ($z=1$) in $d$ dimensional
space. The action of the Einstein-Scalar model is given as
\begin{equation}\label{AdSScalarAction}
  {\cal S}=\frac1{16\pi G_N}\int dx^{d+2} \sqrt{-g}\left( R-\frac12(\nabla \phi)^2-V(\phi) \right),
\end{equation}
whose equations of motion are read as
\begin{subequations}\label{AdSScalarEOM}\begin{align}
  &0= R_{\mu \nu }-\frac1{d}g_{\mu \nu }V(\phi)-\frac12 \partial_\mu \phi \partial_\nu \phi, \\
  &0= \nabla^2\phi - V'(\phi).
\end{align}\end{subequations}

For a constant $\phi_*$ satisfying $V'(\phi_*)=0$ and $V(\phi_*)<0$, there is an $AdS_{d+2}$ solution
\begin{equation}\label{AdSScalarAdS}
  ds^2=\frac{L^2}{r^2}(-dt^2+dr^2+d\textbf{x}^2), \quad -V(\phi_*)L^2=(d+1)d,\quad \phi=\phi_*,
\end{equation}
where $\textbf{x}=(x_1,x_2,\cdots,x_d)$. At finite temperature
the bulk geometry can be described by an AdS-Schwarzschild black
hole with flat horizon
\begin{equation}\label{AdSBHinr}
  ds^2=\frac{L^2}{r^2}\left(- f(r) dt^2 + f(r)^{-1}dr^2 + d\textbf{x}^2\right),\quad f(r)=1-\left(\frac{r}{r_h}\right)^{d+1},
\end{equation}
where $r_h$ is the location of the horizon. We have presented the detailed derivation of the holographic butterfly effect and charge diffusion over a general black hole background in Appendix \ref{AppendixB}. According to the horizon formula (\ref{TFormula}) and (\ref{vBformula}), the butterfly velocity $v_B$ and temperature $T$  are
\begin{equation}\label{vBtemnodeform}
  v_B^2=\frac{d+1}{2d},\quad T=\frac{d+1}{4 \pi r_h}.
\end{equation}

Now we introduce a deformation of the scalar field by turning on its source, which will back-react to the metric. Suppose the potential $V(\phi)$ can be
expanded near $\phi=\phi_*$ as
\begin{equation}\label{ScalarPotential}
  V(\phi)=V(\phi_*) +\frac{m^2}{2}(\phi-\phi_*)^2 + \cdots.
\end{equation}
Then the equation of motion for the scalar field leads to the
following deformation uniformly
\begin{equation}\label{ScalarExpandr}
  \phi=\phi_* +\phi_- r^{\Delta_-}+\cdots +\phi_+ r^{\Delta_+}+\cdots,
\end{equation}
where $\Delta_\pm=\frac12\left(d+1\pm\sqrt{(d+1)^2+4
m^2L^2}\right)$. For later convenience, we define a parameter
$\vartheta\equiv\frac{\Delta_-}{d+1}$. As the calculation
presented in Appendix \ref{AppendixAdSAna}, the scalar field
equation at $O(\phi-\phi_*)$ give
\begin{equation}\label{Greenfunction}
\phi_+=\phi_-H\left(\vartheta\right)r_h^{\Delta_--\Delta_+}+O(\phi_-^2),
\end{equation}
where the specific expression of function $H(\vartheta)$ is
given in (\ref{Hexpression}). Such a relation is
nothing but the Green function of $\cal O_\phi$ at the conformal
fixed point once the quantization method is specified. The scalar field
back-reacts to the metric at order $O((\phi-\phi_*)^2)$, and
affects $v_B$ through the horizon formula (\ref{vBformula}).
We are mainly concerned with the relevant or `weakly'
irrelevant deformation, which is subject to the condition $-\frac12<\vartheta<\frac12$.
The result is
\begin{equation}\label{AdSScalarvB2r}
  v_B^2 = \frac{d+1}{2d} \left( 1 - \phi_-^2 r_h^{2\Delta_-} I\left(\vartheta\right) \frac{d+1}{2d} \right)+ O(\phi_-^3),
\end{equation}
where the specific expression of function $I(\vartheta)$ is
given in (\ref{integral}). When
$-\frac12<\vartheta<\frac12$, we find $I(\vartheta)\geq0$ with
the equality only for $\vartheta=0$. It tells us
that $v_B$ always decreases up to $O(\phi_-^2)$ except for a
marginal deformation. It seems this result coincides with the
argument presented in \cite{Feng:2017wvc}, where it is
found that AdS-Schwarzschild black hole has the maximal
$v_B$ in Einstein-Scalar model (\ref{AdSScalarAction}) when
entropy is fixed. Nevertheless, we point out in Appendix
\ref{AppendixVthVE} that one of their hypotheses in
\cite{Feng:2017wvc} is actually violated in our model, following the perturbation
analysis presented in Appendix \ref{AppendixAdSAna}.

According to (\ref{VBscale}), we expect that the second term in
(\ref{AdSScalarvB2r}) can be expressed in terms of the
dimensionless source $\kappa$ and the other scale $T$. Firstly, we point out that $T$ is still related to $r_h$ by
(\ref{vBtemnodeform}) up to $O(\phi_-)$, since
the scalar field does not back-react to the metric at
$O(\phi-\phi_*)$. To ensure this, we obtain the variation of $T$ at
$O(\phi_-^2)$ for $0<\vartheta<\frac12$ in (\ref{AdSScalarT}), indicating that it is not important enough to correct
$v_B$ at $O(\phi_-^2)$, indeed. Secondly, we remark that the identification of $\kappa$ depends on
the choice of quantization method as well as the form
of the deformation $W(\cal O_\phi)$. We present our specific consideration as follows.

We first focus on the standard quantization with $[{\cal
O}_\phi]=\Delta_+$. According to the asymptotic expansion
(\ref{ScalarExpandr}), the source $\kappa_s$ of $W(\cal O_\phi)$
is identified as \cite{Witten:2001ua}
\begin{equation}\label{sourceSta}
  \phi_-=\kappa_s W'\left((\Delta_+-\Delta_-)\phi_+ \right),
\end{equation}
where the subscript `$s$' of $\kappa_s$ denotes the standard
quantization.

For single trace deformation  $W(\cal O_\phi)=\cal O_\phi$, we
have $1/\nu=[\kappa_s]=\Delta_-$. Such deformation is relevant.
From (\ref{sourceSta}), we identify $\kappa_s=\phi_-$ and obtain
\begin{equation}\label{vBdeform}
  v_B^2 = \frac{d+1}{2d} \left( 1 - \gamma \frac{\kappa_s^2}{ T^{2/\nu}} \right)+ O(\kappa_s^3),
\end{equation}
where
$\gamma=\frac{(d+1)I\left(\vartheta\right)}{2d}\left(\frac{d+1}{4\pi}\right)^\frac2\nu$.
We check our analytical result in (\ref{vBdeform}) by
numerically constructing an AdS domain wall at finite temperature
in Appendix \ref{AppendixAdSNum}. Both relevant and weakly
irrelevant cases are well testified.

For double trace deformation $W({\cal O}_\phi)=\frac12{\cal
O}_\phi^2$, we have $[{\cal O}_\phi^2]=2\Delta_+>d+1$ and
$1/\nu=[\kappa_s]=\Delta_--\Delta_+<0$. Such deformation is
irrelevant. $\phi_-$ is offset in (\ref{sourceSta}) and a critical
temperature is obtained as \cite{Faulkner:2010gj}
\begin{equation}\label{DoubleTraceSta}
  T_c=\frac{d+1}{4\pi}\left(-\frac{H\left(\vartheta\right)\kappa_s}{\nu}\right)^\nu.
\end{equation}
We will see later that the physical meaning of $T_c$ becomes more
transparent in the alternative quantization.

If $\frac12-\frac1{d+1}<\vartheta<\frac12$, i. e.
$\frac{d+1}{2}-1<\Delta_-<\frac{d+1}{2}$, we can choose
alternative quantization with $[{\cal O_\phi}]=\Delta_-$.
The source $\kappa_a$ of $W(\cal O_\phi)$ is identified by
\begin{equation}\label{sourceAlt}
  (\Delta_--\Delta_+)\phi_+=\kappa_a W'\left(\phi_- \right),
\end{equation}
where the subscript `$a$' of $\kappa_a$ denotes alternative quantization.

For single trace deformation $W(\cal O_\phi)=\cal O_\phi$, we have $1/\nu=[\kappa_a]=\Delta_+$. Such deformation is relevant. We identify $\kappa_a=\phi_+$ and obtain
\begin{equation}
  v_B^2 = \frac{d+1}{2d} \left( 1 - \gamma \frac{\kappa_a^2}{ T^{2/\nu}} \right)+ O(\kappa_a^3),
\end{equation}
where $\gamma=\frac{(d+1)I(\vartheta)}{2d H(\vartheta)}\left(\frac{d+1}{4\pi}\right)^\frac2\nu$.

For double trace deformation $W({\cal O}_\phi)=\frac12{\cal
O}_\phi^2$, we have $[{\cal O}_\phi^2]=2\Delta_-<d+1$ and
$1/\nu=[\kappa_a]=\Delta_+-\Delta_->0$. Such deformation is
relevant. Similar to the case of standard quantization, there is a
critical temperature
\begin{equation}\label{DoubleTraceAlt}
  T_c=\frac{d+1}{4\pi}\left(-\frac{\nu\kappa_a}{H\left(\vartheta\right)}\right)^\nu.
\end{equation}
(\ref{DoubleTraceAlt}) and (\ref{DoubleTraceSta}) become
identical as $\kappa_a=-1/\kappa_s$, which is obtained just by
exchanging the source and the expectation value\footnote{It
should be cautious that the definitions of $\nu$ in
(\ref{DoubleTraceAlt}) and (\ref{DoubleTraceSta}) are different.}.
If $\vartheta>0$, $H<0$ always, then $T_c$ is well defined
only when $\kappa_a<0$. When $\kappa_a>0$, double trace
deformation does not affect the metric and $v_B$, at least
at $O(\kappa_a^2)$. It is not surprising since such
 deformation with $\kappa_a>0$ drives a RG flow from the UV fixed point
with $[{\cal O}_\phi]=\Delta_-$ to the IR fixed point with $[{\cal
O}_\phi]=\Delta_+$  \cite{Witten:2001ua}, but does not
affect the geometry classically \cite{Gubser:2002zh}. To
investigate the possible effects of such flow on $v_B$, one
should further study quantum corrections. However, when one attempts to go
to the subleading order with finite $N$, the first term $\frac{d+1}{2d}$ of $v_B$ may
receive corrections as well, which makes the effect of such flow
unclear. When $\kappa_a>0$, Faulkner {\it et. al.}
\cite{Faulkner:2010gj,Faulkner:2010fh} found that a new
instability at $T\leq T_c$, where the scalar field
will condensate with a mean-field critical exponent at finite
temperature. Then $v_B$ will behave just like undergoing the phase transition in a holographic superconductor model, whose derivative is
discontinuous at the phase transition point \cite{Ling:2016wuy}.

Next we turn to investigate the charge diffusion bound (\ref{Diffusionscale}) in this holographic model.
We introduce an electromagnetic term into the action and then consider the perturbations of electromagnetic field $A$
\begin{equation}\label{ElectroS}
  {\cal S}_c=\frac1{16\pi G_N}\int dx^{d+2} \sqrt{-g}(-\frac{1}{4} F^2)
\end{equation}
where $F$ is the field strength
$F=dA$. We calculate the change of $T$ and $D_c$ for $\vartheta>0$ in Appendix
\ref{AppendixAdSAna} and then substitute them into the
diffusion ratio (\ref{Diffusionscale}). For the case of standard
quantization and single trace deformation, the result is
\begin{equation}\label{Dcdeform}
  \frac{D_c\lambda_L}{v_B^2}=\Psi(0)\left(1+\eta\frac{\kappa_s^2}{ T^{2/\nu}}\right) + O(\kappa_s^3),
\end{equation}
where
\begin{equation}
  \eta = J\left(\vartheta,d\right)\left(\frac{d+1}{4\pi}\right)^\frac2\nu, \quad
  \Psi(0)= \left\{\begin{array}{ll}
  \log \left(\frac{\Lambda_\text{UV}}{2\pi T} \right), & \text{ for } d=1 \\
  \frac{d}{d-1}, & \text{ for } d>1
\end{array}\right.
\end{equation}
and $\Lambda_\text{UV}$ is the UV cutoff, which reflects the UV
sensitivity of $D_c$ at $d=1$ \cite{Blake:2016wvh}. The expression
of $J(\vartheta,d)$ is given in (\ref{DcovB2taud1}) or (\ref{DcovB2tau}).
We have numerically estimated $J(\vartheta,d)$ for a wide
range of parameters $(\vartheta,d)$ and found that it is always
non-negative. It means that when the system goes away from
the quantum critical region, the bound is more solid.

Finally, we may investigate
(\ref{OriginalDiffusionBound}) for the momentum transport. In a
large, neutral and homogeneous system, according to the
thermodynamic relation \cite{Gubser:2009cg,Caldarelli:2016nni}
\begin{equation}\label{ThermodynamicRelation}
\epsilon=Ts-p,
\end{equation}
the momentum diffusion constant is \cite{Blake:2016wvh}
\begin{equation}
  D_p=\frac{\eta}{\epsilon+p}=\frac{\eta}{Ts}=\frac1{4\pi T},
\end{equation}
where $\epsilon$ is the energy density and $s$ is the
entropy density, while $p$ is the pressure. The saturated
KSS bound $\eta/s=1/(4\pi)$ in Einstein gravity is also used. Then
its diffusion ratio is
\begin{equation}\label{Dpdeform}
  \frac{D_p \lambda_L}{v_B^2}=\frac1 {2v_B^2},
\end{equation}
which increases as well when the system goes away from the quantum critical region.

\section{Lifshitz black hole with scalar deformation}\label{SectionLif}
In previous section we have analysed $v_B$ and $D_c\lambda_L/v_B^2$ in holographic systems with $AdS_{d+2}$ geometry deformed by scalar field. In this section we show that it can be generalized to the Lifshitz case
\cite{Kachru:2008yh}. Lifshitz spacetime can be found in massive
vector model \cite{Taylor:2015glc,Taylor:2008tg}
\begin{equation}\label{LifScalarAction}
  {\cal S}=\frac1{16\pi G_N}\int dx^{d+2} \sqrt{-g}\left( R-\frac14 {\cal G}^2-\frac12 W {\cal B}^2-\frac12(\nabla \phi)^2-V(\phi) \right),
\end{equation}
where $\cal B$ is a one-form field and ${\cal G}=d {\cal B}$. To
study the scalar deformation, we have added a minimally
coupled scalar field into the action. It allows a Lifshitz solution
\begin{equation}\label{LifSpacetime}\begin{split}
  &ds^2=L^2\left(-\frac{dt^2}{r^{2z}}+\frac{dr^2+d\textbf{x}^2}{r^2}\right), \quad {\cal B}=L\sqrt\frac{2(z-1)}{z} r^{-z} dt, \quad \phi=\phi_*,  \\
  &WL^2=dz ,\quad -V(\phi_*)L^2=z^2+z(d-1)+d^2, \quad V'(\phi_*)=0,
\end{split}\end{equation}
where $z$ is the dynamical critical exponent. This solution
enjoys a scaling symmetry with scaling dimension $[t]=-z,\,
[r]=-1,\, [x]=-1$. It is worthwhile to point out that the
Lifshitz solution with massive vector (\ref{LifSpacetime}) can
smoothly go back to the AdS solution (\ref{AdSScalarAdS}) if
one sets $z=1$, while the Lifshitz solution with running
dilaton can not \cite{Taylor:2015glc,Taylor:2008tg}.

To study butterfly effect, one should construct a bulk
geometry with finite temperature. While, a general analytical
Lifshitz black hole with flat horizon in massive vector model
has not yet been found \cite{Taylor:2015glc}. But we
still expect the scaling dimension of temperature
$[T]=-[t]=z$. For the expansion of potential
$V(\phi)$ in (\ref{ScalarPotential}), the scalar field has the
same expansion as (\ref{ScalarExpandr}) but with different scaling
dimensions $\Delta_\pm=\frac12\left(d+z\pm\sqrt{(d+z)^2+4
m^2L^2}\right)$. The deformation of the scalar back-react to the
metric at $O((\phi-\phi_*)^2)$. According to scaling analysis, we
expect
\begin{equation}\label{LifScalarvB}
  v_B^2=T^{2-\frac2z}\Phi(0)(1-\gamma\frac{\phi_-^2}{T^\frac{2\Delta_-}{z}}) + O(\phi_-^3)
\end{equation}
with two undetermined constants $\Phi(0)$ and $\gamma$. For
the case of single trace deformation, we can identify
$\kappa=\phi_-$ and $1/\nu=\Delta_-$. (\ref{LifScalarvB}) should
go back to (\ref{vBdeform}) if we set $z=1$.

\begin{figure}
  \includegraphics[height=110pt]{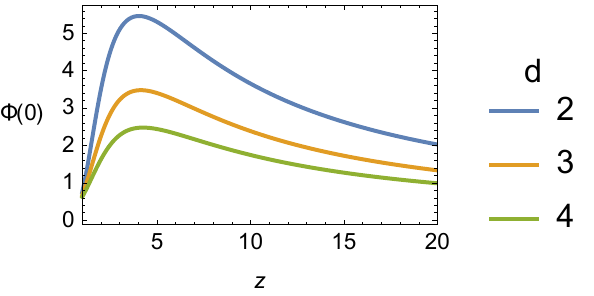}
  \includegraphics[height=110pt]{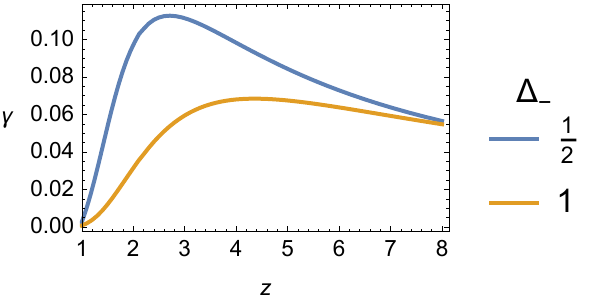}
  \caption{The two coefficients $\Phi(0)$ and $\gamma$ in the expansion (\ref{LifScalarvB}) as functions of dynamical exponent $z$. In the right plot, spatial dimension is $d=2$.}\label{FigLif}
\end{figure}

In Appendix \ref{AppendixLifNum}, we numerically build a Lifshitz
black hole with or without scalar. When scalar field vanishes, we find that $\Phi(0)$ is not equal to
$\frac{d+z}{2d}$, which is the coefficient obtained in Lifshitz
black hole with running dilaton in
\cite{Roberts:2016wdl,Blake:2016wvh}. Such discrepancy is not
surprising, since Lifshitz black hole with running dilaton
is not a solution of (\ref{LifScalarAction}). When scalar field is
turned on, we use standard quantization and impose the boundary
conditions on $\phi_-$. Indeed, our numerical results match
(\ref{LifScalarvB}) at small $\phi_-$ when
$0<\Delta_-<\frac{d+z}{2}$. The coefficients $\Phi(0)$ and
$\gamma$ are shown in Figure \ref{FigLif}. Within our
observation, $\Phi(0)$ and $\gamma$ are always positive.

\section{Comments on other phases at low temperature and phase transitions}\label{SectionOther}
In this section we will focus on the behavior of $v_B$ in
some low temperature phases in the non-quantum-critical region
$T\ll|\kappa|^{z\nu}$. We demonstrate that no universal behavior is observed in these low temperature phases or holographic BKT transition, which is in contrast with the results for quantum critical region controlled by a single fixed point, as we have investigated in previous sections. In next two subsections, a gapless
phase and a gapped phase in the holographic framework with
hyperscaling violation (HV) will be investigated. A relevant
scalar deformation (\ref{ScalarExpandr}) with standard
quantization and single trace deformation can drive the AdS
solution (\ref{AdSScalarAdS}) from the UV region to these HV solutions
in the IR region. We will identify $\kappa=\phi_-$ and
$1/\nu=\Delta_-$, where $\Delta_->0$ is required for a relevant
deformation. In the last subsection, a holographic model
with quantum BKT phase transition is discussed.

\subsection{Gapless phase with hyperscaling violation}
In the Einstein-Scalar model (\ref{AdSScalarAction}), we choose the asymptotic behavior of the potential $V(\phi)$ at $\phi\to\infty$ as
\begin{equation}\label{HVpotential}
  V(\phi\to\infty)\sim V_0 e^{\alpha\phi}(1 + V_1 e^{\alpha_1\phi}),
\end{equation}
in order to construct a AdS-HV domain wall from
$\phi=\phi_*$ to $\phi\to\infty$. The term $V_1$ is common in
the UV completion of HV geometry where the scalar $\phi$
serves as the dilaton
\cite{Kiritsis:2015oxa,Gouteraux:2012yr,Charmousis:2010zz}. We can heat up the
system to finite but low temperature $T\ll|\kappa|^\nu$ by
perturbing a small HV black hole in the IR
\begin{equation}\label{HV}\begin{split}
  ds^2&=L_\text{HV}^2r^{\frac{2\theta}{d}-2}(1+g_1 r^\delta+\cdots) \left[-\left(1-c_T r^{d-\theta+1}+\cdots\right)dt^2+\left(1+c_T r^{d-\theta+1}+\cdots\right)dr^2+d\textbf{x}^2 \right],\\
  e^\phi&=r^\epsilon(1+\phi_1 r^\delta+\cdots),
\end{split}\end{equation}
where
\begin{equation}\label{HVpara}
  \alpha\epsilon=-\frac{2\theta}{d},\quad \epsilon^2=-\frac{2\theta}{d}(d-\theta),\quad \delta=\alpha_1\epsilon,\quad -V_0L_\text{HV}^2=(d-\theta ) (d-\theta +1).
\end{equation}
The region of $0<\theta\leq d+1$ is excluded by the
requirement of relevant thermal mode $c_T$ and Null Energy
Condition \cite{Gouteraux:2012yr}. When $\theta>d+1$, the solution
is found to be gapped and thermodynamically unstable
\cite{Liu:2013una,Kiritsis:2015oxa}. We will discuss this in
the next subsection. In this subsection, we focus on the case of
$\theta<0$, where the solution is found to be gapless and
thermodynamically stable
\cite{Liu:2013una,Kiritsis:2015oxa,Dong:2012se}. The IR is located
at $r\xrightarrow{IR}+\infty$, where the induced line element
vanishes.  There are two perturbation modes in (\ref{HV}), whose
coupling is approximately negligible if they are small enough. The
mode of $\{g_1,\phi_1\}$ is generated by the second term $V_1$ in
(\ref{HVpotential}), where the coefficients $\{g_1,\phi_1\}$ can
be solved in the series expansion about $V_1$. The mode of $c_T$
corresponds to perturbing a small black hole with horizon
$r_h=c_T^\frac1{d-\theta+1}$ and temperature
$T=\frac{|d-\theta+1|}{4\pi r_h}$. We find that above two
modes are most important to the variation of $v_B$. One can
consider other modes in (\ref{HV}), and will find that their
scaling dimensions are $0$ or $d-\theta+1$
\cite{Gouteraux:2012yr}. Except for the thermal mode of $c_T$, all the
relevant modes should not be stimulated for a stable
HV solution in the IR. The marginal modes could be introduced
but they only provide secondary contribution to $v_B$ when compared with two modes in (\ref{HV}).

Plugging (\ref{HV}) into the horizon formula of $v_B$
(\ref{vBformula}), we obtain $v_B$ up to the subleading
order
\begin{equation}\label{HVvB}
  v_B^2\approx\frac{d-\theta+1}{2(d-\theta)} \left(1+ \frac{d \delta}{2 (d- \theta) } g_1 r_h^{\delta }\right)= \frac{d-\theta+1}{2(d-\theta)} \left(1+ \gamma \left(\frac{\kappa^\nu}{T}\right)^\delta\right),\quad T\ll|\kappa|^{z\nu},
\end{equation}
where the approximate equality is used since we neglect the
coupling between these two modes of perturbations. This final
result is obtained based on the following analysis. In the
UV ($r\to0$), $\phi$ is expanded as (\ref{ScalarExpandr}). One can
find the scaling relation $\kappa^\nu=\phi_-^{1/\Delta_-}=Z_g
g_1^{1/\delta}$ where $Z_g$ is a constant which can not be
determined by scaling analysis but relies on the specific
form of the potential $V(\phi)$. Then we obtain the final result
of (\ref{HVvB}) with constant $\gamma=\frac{d \delta}{2 (d-
\theta)} \left(\frac{|d-\theta+1|}{4\pi Z_g}\right)^\delta$.
The power $\delta$ in (\ref{HVvB}) is somehow not a universal
quantity, since it relies on the second exponent $\alpha_1$ of
$V(\phi)$ in (\ref{HVpotential}), which is a tail of the UV
completion process.

\subsection{Gapped phase with hyperscaling violation}
Now we come to the case of $\theta>d+1$. In this case the IR
of (\ref{HV}) is located at $r\xrightarrow{IR}0$. The entropy
density $s$ over the background in (\ref{HV}) behaves as
$s\sim T^{d-\theta}$, leading to a negative specific heat
such that the HV black hole is thermodynamically unstable
\cite{Gursoy:2008za,Kiritsis:2015oxa,Ling:2016ien}. Here we
numerically study the AdS-HV domain wall in $d=2$ with potential
$V(\phi)=-6 -3 \sinh ^2\left(\frac{\phi }{\sqrt{3}}\right)$. In
the UV, the domain wall approaches the AdS spacetimes with
$L=1$, $\Delta_-=1$ and $\Delta_+=2$. While in the IR, it
approaches  the HV geometry with $\theta=8$.

To study the thermodynamics of the system, we should heat it up
and calculate its free energy density $f$. Firstly, from
holographic renormalization
\cite{Skenderis:2002wp,Caldarelli:2016nni}, one notices that the
trace of energy momentum tensor is $-\epsilon+2 p=\langle
T^i{}_i\rangle=\phi_{-}\phi_{+}$ \footnote{An alternative
counterterm associated with the scalar field is proposed in
\cite{Anabalon:2015xvl}, which leads to a different expression of
the trace anomaly.}. Thus, according to the
thermodynamic relation (\ref{ThermodynamicRelation}) and
$f=\epsilon-Ts$, we obtain the expression of free energy density
as \footnote{It also appears in \cite{Astefanesei:2008wz}.}
\begin{equation}\label{FreeEnergy}
f=-\frac13(Ts + \phi_{-}\phi_{+}).
\end{equation}

\begin{figure}
  \includegraphics[height=120pt]{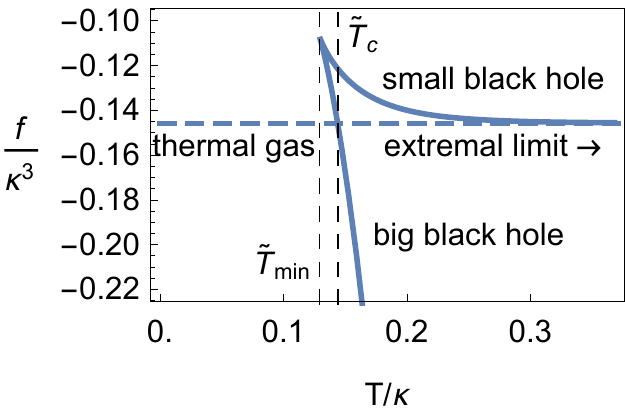} \quad\quad
  \includegraphics[height=120pt]{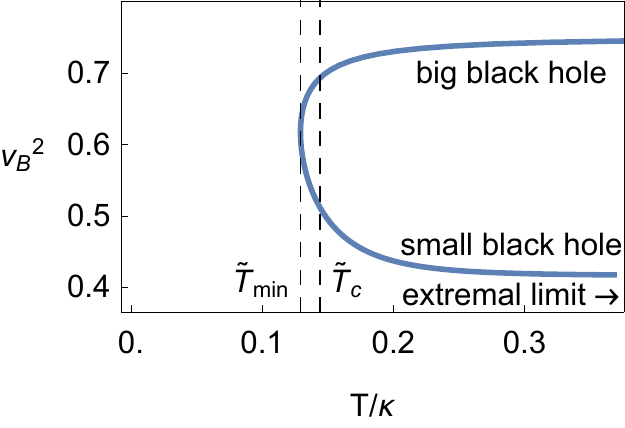}\\
  \caption{Dimensionless free energy density $f/\kappa^2$ and butterfly velocity $v_B^2$ as function of dimensionless temperature $T/\kappa$ for gapped phase with $d=2,\theta=8$. In the left plot, black hole solutions are represented by solid line, thermal gas solution are represented by dashed line.}\label{HVgap}
\end{figure}

We skip the details of numerical analysis since it is similar to the
case of AdS-AdS domain wall as presented in Appendix
\ref{AppendixAdSNum}. We construct dimensionless quantities with the
unit of $\kappa=\phi_{-}$. The temperature dependence of the
dimensionless free energy density $f/\kappa^3$ and the butterfly
velocity $v_B^2$ are shown in Figure \ref{HVgap}. We find two
branches of black hole solutions and a branch of thermal gas
solutions. The branch of big black holes behaves like the
AdS-Schwarzschild black hole (\ref{AdSBHinr}) with positive specific heat and
$v_B^2\leq\frac{d+1}{2d}=\frac34$. While, the branch of small black holes behaves like the HV black hole
(\ref{HV}) with negative specific heat and $v_B^2\geq\frac{d-\theta+1}{2(d-\theta)}=\frac{5}{12}$.
It becomes extremal at $T/\kappa\to\infty$.
There is a minimal dimensionless temperature
$\tilde T_\text{min}$ for those branches of black holes. The branch of
thermal gases has the same form of the extremal solution but
has compact imaginary time $\tau\sim\tau+T^{-1}$
\cite{Gursoy:2008za,Kiritsis:2015oxa}. A critical
temperature $\tilde T_c$ which is higher than $\tilde
T_\text{min}$ appears at the intersection between the
branch of big black holes and the branch of thermal gases in the plot of
free energy density. The thermal gas dominates when
$T/\kappa<\tilde T_c$ while the big black hole dominates
when $T/\kappa>\tilde T_c$. So a first-order phase transition
occurs at $\tilde T_c$.

Specifically, a holographic description of chaos is ill-defined in thermal gas phase since horizon is absent.
So when $T/\kappa$ decreases and the system
falls into such gapped phase with HV, chaos may
disappear and $v_B$ becomes ill-defined when
$T/\kappa<\tilde T_c$.

\subsection{Holographic Berezinskii-Kosterlitz-Thouless transition}\label{SubsectionOtherQCP}

Now we consider a holographic BKT transition which goes beyond the quantum
criticality discussed above, whose QCP is so-called bifurcating
QCP
\cite{Kaplan:2009kr,Iqbal:2010eh,Jensen:2010ga,Iqbal:2011ae,Evans:2010hi,Iqbal:2011aj,Jensen:2010vx}.
From the perspective of RG flow, holographic BKT transition is the result of the annihilation between two fixed points which are linked by the double-trace flow mentioned in section \ref{SectionAdS}.
Consider a scalar field $\phi$ in the bulk and the source of its
dual operator $\cal O_\phi$ in the boundary theory is set to zero.
Consider an $AdS_{n}$ scaling geometry at zero temperature, when
one tunes the effective mass $m_\text{eff}$ of $\phi$ to become
lower than its BF bound $m_\text{BF}$ with respect to
$AdS_{n}$, two fixed points which are respectively dual to
the $AdS_n$ with standard quantization and alternative
quantization will merge and then annihilate \cite{Kaplan:2009kr}.
Then $\phi$ will condense spontaneously and display a
nonzero expectation value $\langle\cal O_\phi \rangle$. The
condensation of $\phi$ generates an intrinsic IR scale
$\Lambda_\text{IR}$ which exhibits a BKT scaling
\cite{Kaplan:2009kr,Jensen:2010ga,Iqbal:2010eh}
\begin{equation}\label{BKTscale}
  \Lambda_\text{IR} \sim \Lambda_{\text{UV},n} e^{-\pi/\tilde\alpha},\quad \tilde\alpha=L_{n}\sqrt{m_\text{BF}^2-m_\text{eff}^2},
\end{equation}
where $\Lambda_{\text{UV},n}$ and $L_{n}$ are the UV cutoff and
the radius of the $AdS_{n}$, respectively. A BKT scaling is
found in the condensation $\langle{\cal O}_\phi \rangle\sim
\Lambda_{\text{UV},n}^\Delta e^{-\pi/(2\tilde\alpha)}$ as well,
where $\Delta$ is the scaling dimension of operator ${\cal
O}_\phi$. So such transition is infinite order and is called
holographic BKT transition. The bifurcating QCP is located
at $\tilde\alpha=0$.

However, when the system goes to finite temperature $T$, the
transition becomes the second order with critical value
$\tilde\alpha_c=\pi \log^{-1}\left(\frac{\Lambda_{\text{UV},n}}{T}\right)$ and mean field exponent $\langle{\cal O}_\phi \rangle\sim \left(\tilde \alpha-\tilde\alpha_c \right)^\frac12 $ \cite{Jensen:2010ga}.

Usually once a bulk action is given, then the mass $m$
of the bulk field $\phi$ is fixed. What we can tune in a QPT is
the source of the operators in the dual QFT. Actually, the
effective mass $m_\text{eff}$ can be tuned by
adjusting the fields which are coupled to $\phi$
\cite{Iqbal:2010eh,Jensen:2010ga,Jensen:2010vx}. For instance, one
can consider a coupling term $\int d^{d+1}x \sqrt{-g}
\frac12Y(\psi)\phi^2$ in the bulk action, where $\psi$ is a
massless axion field and $Y(\psi)$ is a function. The IR
value $\psi_\text{IR}$ of the axion field $\psi$ could be
controlled by the source $\psi_{s}$ of its dual operator. Such
coupling term will contribute to the effective mass
as $m_\text{eff}^2= m^2 - Y(\psi_\text{IR})$. So the
relation between $m_\text{eff}$ and $\psi_s$ depends on
the detail of function $Y(\psi)$. Consequently, by tuning
the source $\psi_s$, one can adjust $m_\text{eff}$ and trigger a BKT
transition according to the mechanism given above. For instance,
some BKT transitions can be triggered by tuning magnetic fields
which lead to condensation of scaler fields on $AdS_2\times R^d$
geometries in the IR \cite{Jensen:2010ga,Evans:2010hi,Jensen:2010vx}.

To further consider the butterfly velocity $v_B$ near the BKT transition, we should study the system at finite temperature. Firstly, we discuss the behavior
of $v_B$ when approaching such bifurcating QCP from the uncondensed phase.
According to the horizon formula of $v_B$ (\ref{vBformula}), one
should investigate the full backreactions of $\psi$ to the
metric. The effects of such backreactions depend on the
potential of $\psi$ and its couplings with other fields.  We expect a
complicated behavior which is different from the simple form
(\ref{AdSScalarvB2r}). Especially, a universal maximization of
$v_B$ near QCP will not appear, since one can easily shift the location of QCP by changing $Y(\psi)$ while leaving the metric as well as the butterfly velocity $v_B$  unchanged in the uncondensed phase.

Next we consider the system undergoes the phase transition and then enters the condensed
phase. Since the phase transition has mean-field scaling at finite temperature,
which is just like a holographic superconductor transition,
we expect that a discontinuity of $\partial v_B/\partial \tilde
\alpha$ will appear at the phase boundary \cite{Ling:2016wuy}.

In addition, when $m_\text{eff}^2>m_\text{BF}^2$, there is a double-trace flow between the fixed
points with standard quantization and alternative quantization. While, as is discussed in
section \ref{SectionAdS}, such flow does not back-react to the
metric and $v_B$ classically.

\section{Comparisons with the results in many-body system}\label{SectionCMT}

In this section we are going to compare our results in holographic approach with the ones obtained
in many body system, including $d=1$ Bose-Hubbard model (BHM) and $d=2$ $O(N)$ nonlinear sigma
model with large $N$
\cite{Shen:2016htm,Bohrdt:2016vhv,Chowdhury:2017jzb}.

\subsection{Bose-Hubbard model}\label{SubsectionBHM}

With the use of numerical simulation, the OTOC (\ref{OTOC})
near the tip of the Mott insulating lob with density $\rho=1$ has
been computed in \cite{Shen:2016htm,Bohrdt:2016vhv}. There is a
Mott insulator-superfluid transition at $U/J\approx3.4$, where $U$
measures the in-site repulsion energy and $J$ measures the
nearest-neighbor hopping energy. Mott insulating phase falls in
the region with $U/J>3.4$ while superfluid phase falls in
$U/J<3.4$. It is known that such kind of transition is a BKT
type which belongs to the universal class of $O(2)$ model in
$d=1$ \cite{Sachdev:1999QPT}. The field theory version of such BKT
transition is sine-Gordon model where the mergence and
annihilation between two fixed points at the QCP is found
\cite{Kaplan:2009kr}. Actually, from the perspective of RG flow,
such BKT transition is controlled by a line of fixed points
rather than a single fixed point \cite{Sachdev:1999QPT}.  A gap
with BKT scaling (\ref{BKTscale}) appears in Mott insulating
phase.

Nevertheless, so far a clear duality between the BKT transition in BHM
and the holographic BKT transition has yet been found. Some difficulties arise when one attempts to compare the
butterfly effect in these two different scenarios. Firstly, the mean-field scaling at finite temperature in holographic BKT is different from that one in the BKT transition of many-body system. Secondly, as discussed in
subsection \ref{SubsectionOtherQCP}, before two fixed points annihilate, the double-trace flow continuously changes. But such change does not
affect the metric at classical level, let alone the butterfly effect. So, as a preliminary approach, we attempt
to provide a phenomenological view on this issue by comparing the results (\ref{vBqcr}) and (\ref{DcovB2tauqcr}) in
the second order QPT with the ones in above Bose-Hubbard model.

\begin{figure}
  \includegraphics[width=150pt]{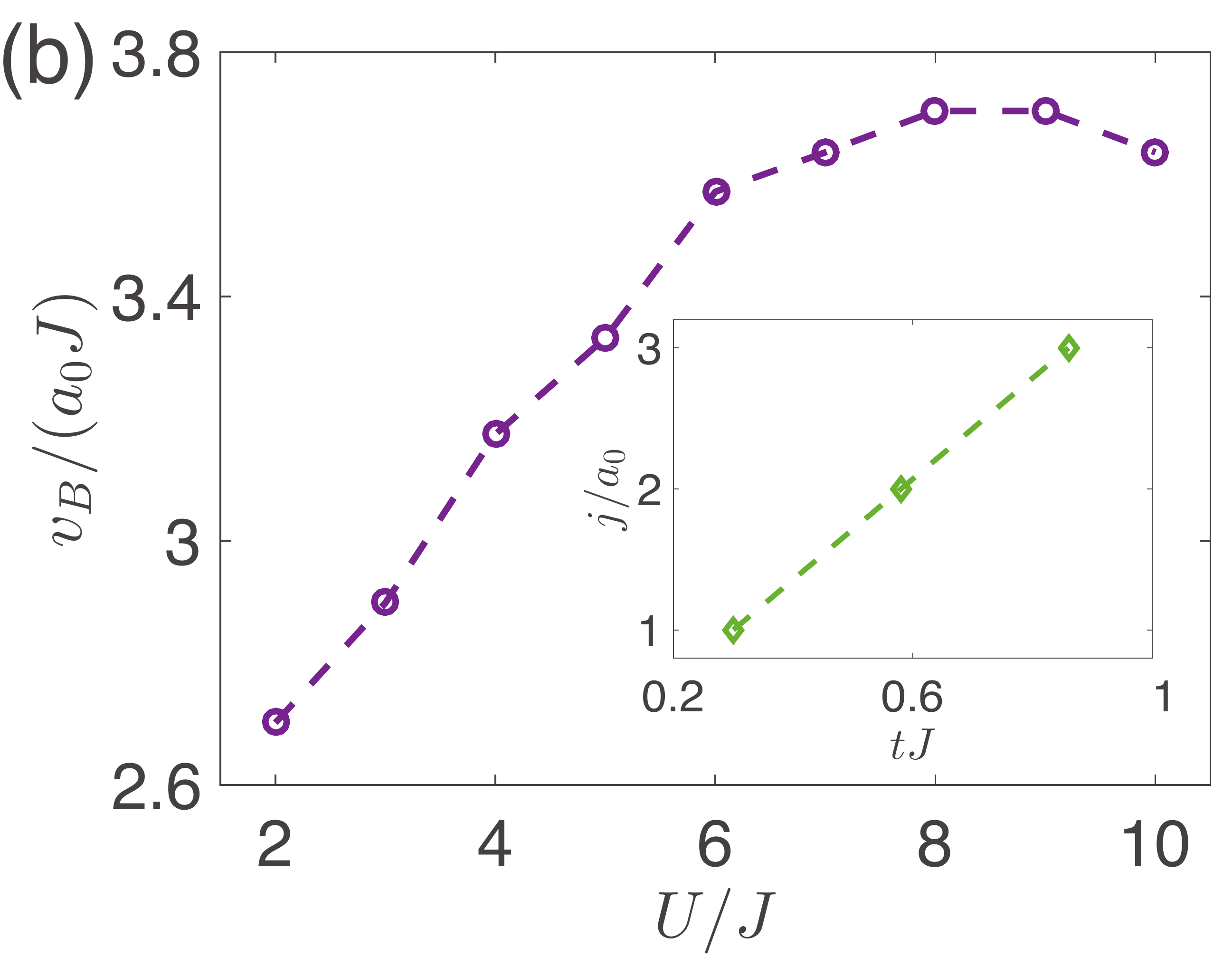}\\
  \caption{Butterfly velocity $v_B$ as a function of coupling $U/J$ at temperature $\beta J=0.9$. This plot is taken from \cite{Shen:2016htm}.}\label{FigOTOCvBShen}
\end{figure}

In \cite{Shen:2016htm}, the authors considered a system with
$7$ sites and $7$ bosons and applied the method of exact
diagonalization. The system goes across the QCP by tuning the
$U/J$ at fixed and finite temperature. The result of butterfly
velocity $v_B$ is shown in Figure \ref{FigOTOCvBShen}, where a
peak of $v_B$ near $U/J\approx9$ is observed. Such value
$U/J\approx9$ is larger than the critical value $U/J\approx3.4$ at
zero temperature. It was conjectured in \cite{Shen:2016htm}
that such deviation may be ascribed to finite temperature and
finite size. In spite of the deviation, the phenomenon found
in this many body system is analogous to our holographic results
(\ref{vBqcr}) with fixed $T$.

\begin{figure}
  \includegraphics[width=150pt]{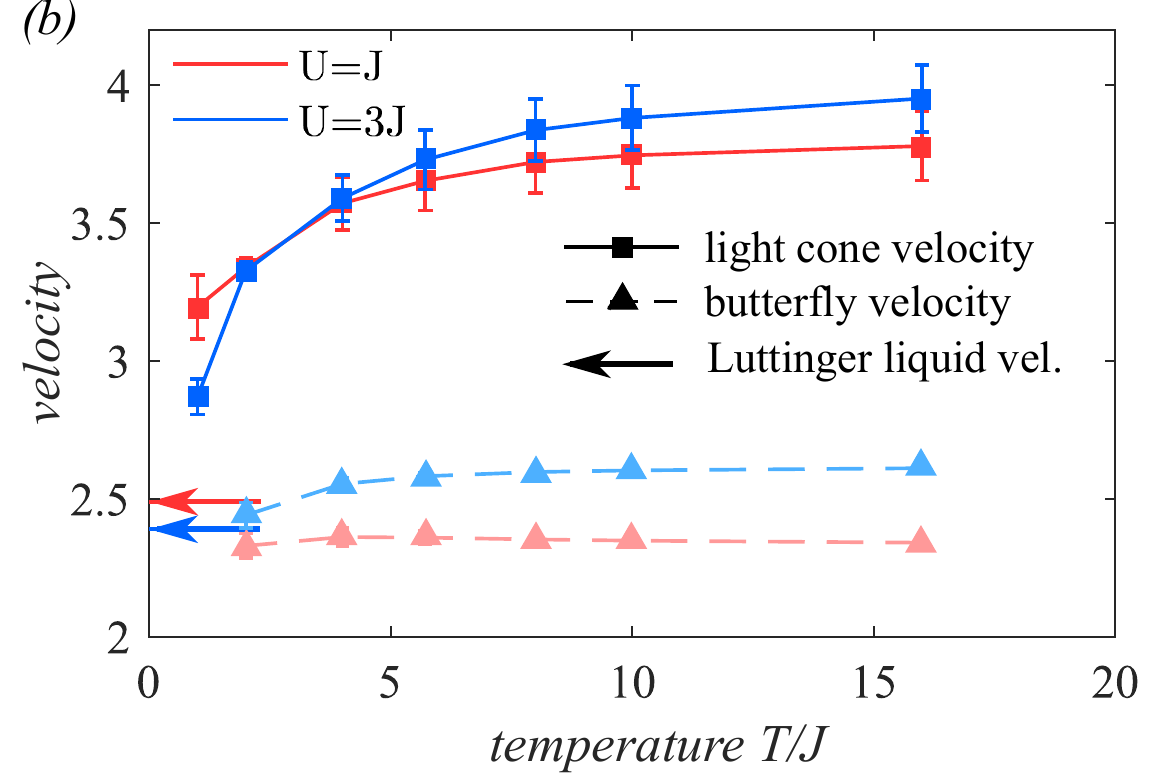}\\
  \caption{Butterfly velocity $v_B$ as a function of temperature $T/J$ for different coupling $U/J$. This plot is taken from \cite{Bohrdt:2016vhv}.}\label{FigOTOCvBBohrdt}
\end{figure}

In \cite{Bohrdt:2016vhv}, the authors perform the numerical simulations
based on matrix-product-operators at finite-temperature. The system size is large enough to guarantee the
convergence. The result of butterfly velocity $v_B$ is shown in
Figure \ref{FigOTOCvBBohrdt}. Let us focus on the case of $U=3J$
where the system is close to QCP at zero temperature. When $T\gg
J$, as shown in Figure
\ref{FigOTOCvBBohrdt}, the system falls in the region of lattice high
temperature and the scaling symmetry does not emerge. When
$T\lesssim J$, the system lies in quantum
critical region. In such region, as $T/J$ goes down, $v_B$
begins to decrease, which coincides with the tendency of our
holographic results (\ref{vBqcr}) with fixed $\kappa$.

\subsection{$O(N)$ nonlinear sigma model}\label{SubsectionONModel}

Let us turn to the $d=2$ $O(N)$ nonlinear sigma model. In
\cite{Chowdhury:2017jzb}, the authors studied the chaos with
QPT up to $1/N$ order. The QCP at $g=g_c$ and $T=0$ is controlled
by a $z=1$ CFT. In the quantum critical region, they focus on
$g=g_c$ and only consider the dominating scale $T$. The inverse
phase coherent time is $\frac1{\tau_\varphi}=1.152 \frac TN$,
which is suppressed by large $N$. Then quasi-particles are still
well defined up to $1/N$ order \cite{Sachdev:1999QPT}. The
dispersion relation at $N=\infty$ is
\begin{equation}\label{DispersionRelation}
  \epsilon_{\bm k}^2=c^2{\bm k}^2+\mu^2,
\end{equation}
where the speed of light $c=1$, the thermal mass $\mu=\Theta T$
and the constant $\Theta=2\log\frac{1+\sqrt5}{2}$. The thermal
mass $\mu$ gives a finite correlation length $\xi=1/\mu$ at
$N=\infty$ \footnote{A holographic thermal screening effect is also
observed in a black hole background \cite{Hartnoll:2016apf}.}.

To study the quantum chaos, the authors in \cite{Chowdhury:2017jzb} consider the interaction up to $1/N$. By calculating the square
commutator (\ref{Commutator}), they extract the
Lyapunov exponent and the butterfly velocity, which are
$\lambda_L\approx 3.2 \frac TN$ and $v_B\approx c$. The Lyapunov
exponent $\lambda_L$ is suppressed by large $N$ and the butterfly
velocity $v_B$ is closed to the specific velocity of
quasi-particle. Both of $\lambda_L$ and $v_B$ are different from
the ones in gravity. It is reasonable, since the dual field theory
of Einstein gravity is considered as large $N$ gauge field
theories at strong coupling.

To discuss the behavior of $v_B$ in the quantum critical region,
as what is done in above sections, we should deviate the system from the QCP
and consider another scale $g-g_c$. It generates the energy gap
$\Delta_E^\pm$, where $\Delta_E^-\sim (g_c-g)^\nu$ for $g\leq g_c$
and $\Delta_E^+\sim (g-g_c)^\nu$ for $g\geq g_c$ and the critical
exponent $\nu=1$. Remind that $\kappa\sim g-g_c$, we have
$\Delta_E^\pm\sim |\kappa|^\nu$ as usual. The energy gap
$\Delta_E^\pm$ only affects the thermal mass $\mu$ with the
scaling formulas at $N=\infty$ \cite{Sachdev:1999QPT}
\begin{equation}\label{ThermaldMass}
  \mu= \left\{\begin{array}{lll}
  2 T \sinh^{-1}\left[\frac12\exp(-2\pi\Delta_E^-/T)\right] &\text{ for }& g\leq g_c \\
  2 T \sinh^{-1}\left[\frac12 \exp(\Delta_E^+/2T)\right] &\text{ for }& g\geq g_c
  \end{array}\right.,
\end{equation}
both of which match $\mu=\Theta T$ at $g=g_c$. From
(\ref{ThermaldMass}), $\mu/T$ monotonously decreases when $g$
increases. So the point $g=g_c$ is not an extreme point of
$\mu/T$. The energy gap $\Delta_E^\pm$ does not explicitly enter the
calculations of $\tau_\varphi,\lambda_L$ and $v_B$ in
\cite{Chowdhury:2017jzb}, except changing the thermal mass
$\mu$. The value of $\mu/T=\Theta=2\log\frac{1+\sqrt5}{2}$ at
$g=g_c$ is not special in the calculations of
$\tau_\varphi,\lambda_L$ and $v_B$, so we do not expect that any
extremal behavior of $\tau_\varphi,\lambda_L$ and $v_B$
would appear at $g=g_c$.

In fact, in the quantum critical region, the saturation of the bound for $\tau_\varphi$ in (\ref{coherencetime})
does not guarantee the minimization of $\tau_\varphi$ at
$g=g_c$. If the absence of extreme for $v_B$ at $g=g_c$ in such model is actually true, it is different
from the result of our holographic calculation. It means that the extreme for $v_B$ may only appear in the model which has quasi-particle description and weakly chaotic behavior. So we expect to find universal behavior of $v_B$ in strong chaotic systems. The complicated
dependence of (\ref{ThermaldMass}) on two scales $T$ and
$\Delta_E^\pm$ implies that, in a many-body system, the
 finite temperature effects in quantum critical region may not be so
simple as that in a holographic system with black holes deformed by a scalar field.

\section{Discussion}

In this paper, we have investigated the butterfly velocity
$v_B$ and the diffusion ratios $D_{c,p}\lambda_L/v_B^2$ near the quantum
phase transition in holographic approach. In the quantum critical region, when the relevant scalar deformation is turned on, butterfly velocity $v_B$ universally decreases such that a local peak of $v_B$ is observed near QCP, whose universal behavior depends on the critical exponents and the dimension of the system. In addition, the diffusion ratios $D_{c,p}\lambda_L/v_B^2$ universally increase, which satisfies the diffusion bound (\ref{OriginalDiffusionBound}). We have also studied the behavior of $v_B$ in low temperature
phases beside the QCP, which is controlled by the
IR fixed points. For the gapless phase with running dilaton and
hyperscaling violation, the variation of $v_B$ mainly relies on
the UV completion process, which is not intrinsic. For the
gapped phase, black hole does not dominate at low
temperature and chaos disappears. In the holographic BKT
transition, the behavior of $v_B$ relies on the
details of the coupling term in the bulk.

The numerical results in BHM support our holographic observation
that $v_B$ decreases when the system goes away from $g=g_c$ but
still within the quantum critical region. So we expect that this
decreasing behavior of $v_B$ is not an occasional phenomenon.

It is instructive to understand the
universality of $v_B$ in quantum critical region from a theoretical point of view. On gravity
side, it seems that one may link it to the well-known holographic
C-theorem \cite{Freedman:1999gp} \footnote{Some correspondences between butterfly velocities and central charges are observed in massive gravity theories \cite{Qaemmaqami:2017jxz}.}. But their connections actually
are not evident. Firstly, central charge, which is related to
the AdS radius $L$, does not appear in the expression of $v_B$.
Secondly, $v_B$ does not monotonously decrease when deformed away from the QCP in the whole
phase diagram of QPT. As illustrated in the context of the AdS-AdS domain wall, $v_B$ decreases firstly and then increases again as
$T/\kappa$ decreases, where the low temperature phase is gapless and controlled by
another CFT. Finally, the value of $v_B$ at low
$T/\kappa$ is the same as the one at high $T/\kappa$.
Similar phenomenon is observed in the $O(N)$ nonlinear sigma model
in \cite{Chowdhury:2017jzb}, where the values of $v_B$ in the
quantum critical region with $g=g_c$ and in the symmetry broken
phase are the same, although the crossover behavior has not been
obtained. Another possible reason for the decrease of $v_B$ in
quantum critical region is the reverse isoperimetric inequality,
which is proposed to be linked to a maximum of $v_B$ in
\cite{Feng:2017wvc}. In Appendix \ref{AppendixVthVE}, we find that
such inequality is true in our perturbation analysis.

On field theory side, the similarity between (\ref{vqp}) and (\ref{vBqcr}) hints the mechanism that the energy gap $\Delta_E$ hinders the spread of the chaos even in a system without quasi-particle. While, the discussion about the $O(N)$ nonlinear sigma model
tells us that such decrease of $v_B$ may not be always true when quasi-particle description is still valid and the chaos is suppressed. Based on the calculation in
\cite{Chowdhury:2017jzb}, the coupling constant $\kappa\sim g-g_c$
affects the chaos only through the thermal mass $\mu$. So we
expect a detailed discussion about the dependence of chaos
on general $\mu$.

Essentially, the BKT transition in BHM is controlled by a fixed line rather than a
single fixed point with relevant deformation. The butterfly effect
in other QPTs deserve to be further studied, such as the Mott
insulator-superfluid transition with fixed density in BHM at $d\geq2$. Perhaps one more direct way of studying $v_B$
under deformation lies on field theory side, such as a generalized
SYK model with relevant deformation
\cite{Gu:2016oyy,Stanford:2015owe,Roberts:2014ifa,Lucas:2017dqa}.

Finally, we remark that for simplicity only scalar deformation is considered in this
paper. It directly leads to a $\kappa^2$-variation of $v_B$ and
$D_{c,p}\lambda_L/v_B^2$ for single trace deformation, since scalar
field usually back-reacts to the metric at second order. It is desirable to explore the possible new features of holographic butterfly effect by considering other kinds of deformations in future. Furthermore, in this paper all the holographic setup is considered only at the classical level, which is dual to a gauge field
theory with $N=\infty$. When the subleading order with $1/N$ corrections is taken into account, it is expected that the dependent behavior of $v_B$ on $\kappa^2$ would receive corrections as well. Last but not least, only Einstein gravity with minimally coupled scalar field is considered in this paper. More general coupling terms such as $f(\phi)R$ or higher order curvature corrections deserve further investigations \cite{Roberts:2014isa,Alishahiha:2016cjk,Qaemmaqami:2017bdn}.

\begin{acknowledgments}

We are very grateful to Yidian Chen, Pengfei Zhang, Ruihua Fan,
Hui Zhai, Xing-Hui Feng, Peng Liu, Wei-Jia Li, Chao Niu, Shaofeng Wu, Xiangrong
Zheng, Mohammad M. Qaemmaqami, Xiao-Xiong Zeng, Matteo Baggioli, Walter Tangarife, Dumitru Astefanesei, Annabelle Bohrdt and Ali Naseh for helpful discussions and correspondence. This work is
supported by the Natural Science Foundation of China under Grant
Nos.11275208 and 11575195. Y.L. also acknowledges the support from
Jiangxi young scientists (JingGang Star) program and 555 talent
project of Jiangxi Province.

\end{acknowledgments}

\begin{appendix}

\section{Scalar deformation on perturbation}\label{AppendixAdSAna}
In this appendix we will consider the scalar perturbation
over the AdS-Schwarzschild black hole (\ref{AdSBHinr}) with the
action (\ref{AdSScalarAction}). We start with the ansatz for the
metric
\begin{equation}\label{AdSScalarAnsatz}
  ds^2=-E(r)dt^2+B(r)dr^2+C(r)d\textbf{x}^2,\quad \phi=\phi(r).
\end{equation}
Then we obtain the equations of motion and zero-energy
constraint
\begin{subequations}\label{AdSScalarEOMA}\begin{align}
 &0=\frac{\left(E'\right)^2}{4 B E^2}+\frac{B' E'}{4 B^2 E}-\frac{d C' E'}{4 B C E}-\frac{V(\phi )}{d}-\frac{E''}{2 B E}, \\
 &0=\frac{d \left(C'\right)^2}{4 B C^2}+\frac{d B' C'}{4 B^2 C}+\frac{\left(E'\right)^2}{4 B E^2}+\frac{B' E'}{4 B^2 E}-\frac{V(\phi )}{d}-\frac{\left(\phi '\right)^2}{2 B}-\frac{d C''}{2 B C}-\frac{E''}{2 B E}, \\
 &0=\frac{d C' \phi '}{2 B C}+\frac{E' \phi '}{2 B E}-\frac{B' \phi '}{2 B^2}+\frac{\phi ''}{B}-V'(\phi ), \\
 &0=-\frac{d \left(C'\right)^2}{4 B C^2}-\frac{C' E'}{2 B C E}+\frac{\left(C'\right)^2}{4 B C^2}+\frac{\left(\phi '\right)^2}{2 B d}-\frac{V(\phi )}{d}, \label{AdSScalarHam}
\end{align}\end{subequations}
where the derivative of $\{E,B,C,\phi\}$ is with respect to $r$
while the derivative of $V$ is with respect to $\phi$.

For convenience, we take the coordinate transformation
$\zeta=1-f(r)=(r/r_h)^{d+1}$ such that in coordinates system
$\{t,\zeta,\textbf{x}\}$ equation (\ref{AdSBHinr}) can be written
into the form as (\ref{AdSScalarAnsatz}), with components
\begin{equation}
  E(\zeta)=(1-\zeta)C(\zeta)=\frac{L^2}{r_h^2}(1-\zeta)\zeta^{-\frac{2}{d+1}},\quad B(\zeta)=\frac{L^2}{(d+1)^2(1-\zeta)\zeta^2}.
\end{equation}
where the asymptotic boundary is located at $\zeta\to0$ and
the horizon is at $\zeta=1$.

Now we turn on the deformation of scalar field as presented
in (\ref{ScalarExpandr}), then the black hole will be
back-reacted by the scalar field. We write such variation into the
series expansion of $\lambda$
\begin{subequations}\label{AdSScalarSch}\begin{align}
  E(\zeta)&=\frac{L^2}{r_h^2}(1-\zeta)\zeta^{-\frac{2}{d+1}}(1+\lambda E_1(\zeta)+\lambda^2 E_2(\zeta)+\cdots),\\
  B(\zeta)&=\frac{L^2}{(d+1)^2(1-\zeta)\zeta^2}(1+\lambda B_1(\zeta)+\lambda^2 B_2(\zeta)+\cdots)    ,\\
  C(\zeta)&=\frac{L^2}{r_h^2} \zeta^{-\frac{2}{d+1}} (1+\lambda C_1(\zeta)+\lambda^2 C_2(\zeta)+\cdots),\\
  \phi(\zeta)&=\phi_*+\lambda \phi_1(\zeta) + \lambda^2 \phi_2(\zeta) + \cdots,
\end{align}\end{subequations}
with the coordinate relation $\zeta=(r/r_h)^{d+1}$ unchanged.
We require that the location of both boundary and horizon should not
be changed by the deformation at higher orders.
We adopt the gauge
\begin{equation}\label{gauge}
  E_i(\zeta)=C_i(\zeta), \quad \text{for}\quad  i=1,2,\cdots.
\end{equation}

The variation of the scalar field only appears in Einstein
equations at $O(\lambda^2)$. The first order deformation of metric
is
\begin{equation}\label{AdSScalarMetricMode1}
  E_1(\zeta)=C_1(\zeta)= c_1 - \frac{c_2}{(1+d)\zeta},\quad B_1(\zeta)=c_2(\frac12-\frac1\zeta).
\end{equation}
where $c_1$ and $c_2$ are two integral constants. $c_1$
corresponds to rescaling $t$ and $\textbf{x}$ which should be set
to zero to fix the coordinates of the dual field theory. $c_2$
controls an irrelevant mode, which should be set to zero
as well for preserving the $AdS$ in the UV. Then
\begin{equation}\label{AdSScalarMetricMode10}
  E_1(\zeta)=B_1(\zeta)=C_1(\zeta)=0.
\end{equation}
The scalar equation at $O(\lambda)$ is
\begin{equation}\label{AdSScalarSE}
   (1-\zeta) \phi_1'' - \phi_1' + \frac{(1-\vartheta) \vartheta}{\zeta^2} \phi_1=0,
\end{equation}
where $\vartheta=\frac{\Delta_-}{d+1}$ and $\Delta_\pm=\frac12\left(d+1\pm\sqrt{(d+1)^2+4 m^2L^2}\right)$. Here we only study the situation that Breitenlohner-Freedman (BF) bound is satisfied, then $\vartheta<\frac12$. The violation of BF bound leads to a holographic Berezinskii-Kosterlitz-Thouless transition in subsection \ref{SubsectionOtherQCP}. $\phi_1$ should be regular at the horizon. The solution up to a constant factor is
\begin{equation}\label{phi1}
  \phi_1(\zeta)=\zeta^\vartheta \, _2F_1(\vartheta,\vartheta;2 \vartheta;\zeta)-\zeta^{1-\vartheta}\frac{ \Gamma (1-\vartheta)^2 \Gamma (2 \vartheta) }{\Gamma (2-2 \vartheta) \Gamma (\vartheta)^2} \, _2F_1(1-\vartheta,1-\vartheta;2-2 \vartheta;\zeta),
\end{equation}
where $_2F_1(a,b;c;\zeta)$ is the Gaussian hypergeometric function. Near $\zeta\to0$, $\phi_1$ is expanded as
\begin{equation}\label{phiExpand}\begin{split}
  \phi_1(\zeta) \sim \zeta^\vartheta +  \cdots + H(\vartheta) \zeta^{1-\vartheta}+ \cdots
  = \left(\frac{r}{r_h}\right)^{\Delta_-} + \cdots + H(\vartheta) \left(\frac{r}{r_h}\right)^{\Delta_+} + \cdots,
\end{split}\end{equation}
where
\begin{equation}\label{Hexpression}
H(\vartheta)= -\frac{ \Gamma (1-\vartheta)^2 \Gamma (2 \vartheta)}{\Gamma (2-2 \vartheta) \Gamma (\vartheta)^2}.
\end{equation}
The mode led by $\zeta^\vartheta$ is relevant when
$\vartheta>0$ while irrelevant when $\vartheta<0$. The mode
led by $\zeta^{1-\vartheta}$ is always relevant. We have
restored the asymptotic expansion of $\phi_1(\zeta)$ with the
$r$ coordinate in order to display the $r_h$ dependence. By
comparing (\ref{phiExpand}) with (\ref{ScalarExpandr}), we can
identify
$\lambda=\phi_-r_h^{-\Delta_-}=\phi_+H(\vartheta)r_h^{-\Delta_+}$
at $O(\lambda)$ and obtain (\ref{Greenfunction}). At $\zeta=1$,
\begin{equation}
  \phi_1(1)=\frac{2 \pi  \cot (\pi  \vartheta) \Gamma (2 \vartheta)}{\Gamma (\vartheta)^2}.
\end{equation}

\begin{figure}
  \includegraphics[width=150pt]{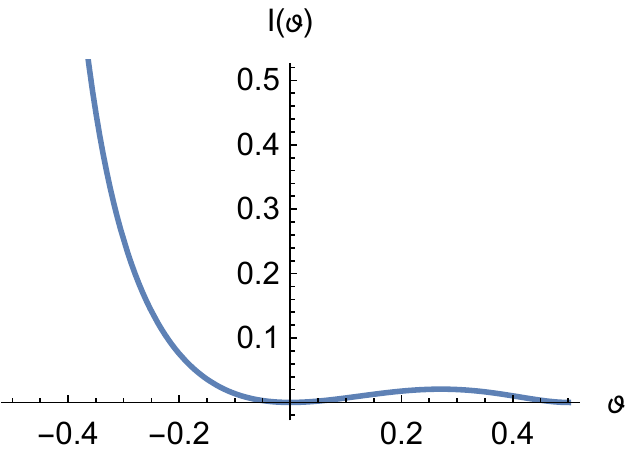}\\
  \caption{$I(\vartheta)$ as a function of $\vartheta$.}\label{Figavsc}
\end{figure}

Now we consider the perturbations at the subleading order
$O(\lambda^2)$. The Einstein equations at $O(\lambda^2)$ are
differential equations for metric $E_2,B_2,C_2$ with source
$\phi_1$ where $\phi_2$ is absent. The deformation of metric is
found to be
\begin{equation}\label{AdSScalarMetricMode2}
  C_2'(\zeta)=E_2'(\zeta)=\frac1{d+1}B_2'(\zeta)=\frac1{d \zeta^2}\left(\int_\zeta^1 y^2 \phi'_1(y)^2 \, dy-c_3\right).
\end{equation}
Similar to the step in (\ref{AdSScalarMetricMode1}), we
demand that the boundary mode $B_2\sim\#\zeta^{-1}$
vanishes, which determines the constant $c_3$ to
be
\begin{equation}\label{integral}
  I(\vartheta)=\int_0^1 y^2 \phi'_1(y)^2 \, dy,
\end{equation}
which is a function of $\vartheta$ and plotted in Figure
\ref{Figavsc}. According to the asymptotic expansion
(\ref{phiExpand}), integral $I(\vartheta)$ diverges when
$\vartheta<-\frac12$, which makes such cancellation subtle.
However, here we only consider the case of relevant or `weakly'
irrelevant deformation. So we assume $-\frac12<\vartheta<\frac12$
from now on. By setting $c_3=I(\vartheta)$, we obtain
\begin{equation}\label{AdSScalarMetricMode20}
  C_2'(\zeta)=E_2'(\zeta)=\frac1{d+1}B_2'(\zeta)=-\frac1{d \zeta^2} I_1(\zeta;\vartheta),
\end{equation}
where $I_1(\zeta;\vartheta)=\int_0^\zeta y^2 \phi'_1(y)^2 \,
dy\sim \frac{\vartheta^2}{2\vartheta+1}\zeta^{2\vartheta+1}$ near
$\zeta\to0$. Obviously, $I(\vartheta)=I_1(1;\vartheta)$.

Now, we can calculate the variation of $v_B$ according to
(\ref{vBformula})\footnote{It should be cautious that the
coordinate $r$ in (\ref{vBformula}) is $\zeta$ now.}. By using
(\ref{gauge}) (\ref{AdSScalarMetricMode10}) and
(\ref{AdSScalarMetricMode20}), it is enough to derive it as
\begin{equation}\label{AdSScalarvB2}
  v_B^2= \frac{d+1}{2d} \left( 1 - \lambda^2 I(\vartheta)\frac{(d+1)}{2 d}\right)+ O(\lambda^3),
\end{equation}
which is (\ref{AdSScalarvB2r}).

On the asymptotic boundary, $\zeta\to0$, the deformation of the
metric behaves as $\zeta^{2\vartheta}$, which is
vanishing when $0<\vartheta<\frac12$ and divergent
when $\vartheta<0$. From now on, we only discuss the case of
$0<\vartheta<\frac12$. By applying zero-energy constraint
(\ref{AdSScalarHam}) at $O(\lambda^2)$ on $\zeta\to 0$ and requiring that constant modes of $C_2,E_2,B_2$ vanish, we can
obtain
\begin{equation}\label{AdSScalarMetricMode21}
  C_2(\zeta)=E_2(\zeta)=\frac1{d+1}B_2(\zeta)=-\frac1{d} \int_0^\zeta y^{-2} I_1(y;\vartheta) dy\equiv -\frac1{d} I_2(\zeta;\vartheta).
\end{equation}

According to (\ref{TFormula}), the Hawking temperature $T$
is
\begin{equation}\label{AdSScalarT}
   T    = \frac{d+1}{4 \pi  r_h} \left(1+ \frac{\lambda^2}2 I_2(1;\vartheta) \right) +O(\lambda^3).
\end{equation}
By combining (\ref{AdSScalarvB2}), (\ref{TFormula}) and (\ref{LyapunovExp}), we obtain
\begin{equation}
  \frac{v_B^2}{\lambda_L}
  =\frac{r_h}{d}-\frac{\lambda ^2 r_h ((d+1) I_1(\vartheta,1)+d I_2(\vartheta,1))}{2
  d^2}.
\end{equation}

Charge diffusion constant $D_c$ can be evaluated from
(\ref{Dcformula}). For $d=1$, it depends on the UV cutoff
$\Lambda_\text{UV}$ in the original $r$ coordinate
\begin{equation}
  D_c=r_h \log \left(\Lambda_\text{UV}  r_h\right) \left(1+\frac{1}{2} \lambda ^2 \left(I_2(\vartheta,1)-\frac{I_3(\vartheta,-1)}{\log \left(\Lambda_\text{UV}  r_h\right)}\right)\right).
\end{equation}
For $d>1$, the result is
\begin{equation}\begin{split}
  D_c
  = \frac{r_h}{d-1} + \frac{\lambda ^2 r_h}{2 d \left(d^2-1\right)} \left(\left(-d^2+d+2\right) I_2(1;\vartheta)-2 (d-1) I_3\left(\vartheta,-\frac{2}{d+1}\right)\right),
\end{split}\end{equation}
where $I_3(\vartheta,p)=\int_0^1 y^p I_2(y;\vartheta) dy$.

Finally, one can calculate the diffusion ratio $D_c\lambda_L/v_B^2$. For $d=1$,
\begin{equation}\label{DcovB2taud1}\begin{split}
  \frac{D_c\lambda_L}{v_B^2}&=\log \left(\frac{\Lambda_\text{UV}}{2\pi T} \right)\left(1+\lambda^2 J(\vartheta;1)\right),  \\
  J(\vartheta;1)&=I_1(1;\vartheta)+I_2(1;\vartheta)+\frac{I_2(1;\vartheta)-I_3(\vartheta,-1)}{2 \log \left(\frac{\Lambda_\text{UV}}{2\pi T} \right)},
\end{split}\end{equation}
Since $\Lambda_\text{UV}\gg T$, the logarithmic terms $\log \left(\frac{\Lambda_\text{UV}}{2\pi T} \right)$ is usually large such that the third term of $J(\vartheta;1)$ is negligible.
For $d>1$,
\begin{equation}\label{DcovB2tau}\begin{split}
  \frac{D_c\lambda_L}{v_B^2}&=\frac{d}{d-1}\left(1+\lambda^2 J(\vartheta,d)\right)  ,\\
  J(\vartheta,d)&=\frac{(d+1)^2 I_1(1;\vartheta)+2 (d+1) I_2(1;\vartheta)-2 (d-1) I_3\left(\vartheta,-\frac{2}{d+1}\right)}{2 d (d+1)}.
\end{split}\end{equation}
We numerically evaluate $J(\vartheta,d)$ for a wide range of
$\{\vartheta,d\}$ and find it is always non-negative. Those
integrals are collected here
\begin{equation}\label{I123}\begin{split}
  I(\vartheta)&=\int_0^1 y^2 \phi'_1(y)^2 \, dy   ,\\
  I_1(\zeta;\vartheta)&=\int_0^\zeta y^2 \phi'_1(y)^2 \, dy \sim \frac{\vartheta^2}{2\vartheta+1} \zeta^{2\vartheta+1}  ,\\
  I_2(\zeta;\vartheta)&=\int_0^\zeta y^{-2} I_1(y;\vartheta) dy \sim \frac{\vartheta}{2(2\vartheta+1)} \zeta^{2\vartheta} ,\\
  I_3(\vartheta,p)    &=\int_0^1 y^p I_2(y;\vartheta) dy    \sim \frac{\vartheta}{2(2\vartheta+1)(2\vartheta+p+1)} \quad \text{if}\quad  2\vartheta+p+1>0.
\end{split}\end{equation}

\section{Formula of butterfly velocity and charge diffusion constant}\label{AppendixB}
In this appendix we derive the formulas of $v_B$ and $D_c$,
closely following the strategy presented in
\cite{Roberts:2016wdl,Blake:2016wvh,Roberts:2014isa}. Given
a black hole metric in $d+2$ dimensions with the form
\begin{equation}
  ds^2=-E(r)dt^2+B(r)dr^2+C(r)d\textbf{x}^2,
\end{equation}
whose components are expanded near the horizon $r_h$ as
\begin{equation}
  E(r)\sim E'(r_h)(r-r_h)+\cdots,\quad B(r)\sim B^{(-1)}(r_h)(r-r_h)^{-1}+\cdots,\quad C(r)\sim C(r_h)+C'(r_h)(r-r_h)+\cdots,
\end{equation}
where $E'(r_h),B^{(-1)}(r_h),C'(r_h)$ are finite and negative. We
consider flat horizon and require that the asymptotic boundary is
located at $r\to0$. Then the black hole temperature and entropy
density are
\begin{equation}\label{TFormula}\begin{split}
  T&=\frac1\beta=\frac{-E'(r_h)}{4\pi\sqrt{B(r_h)E(r_h)}}=\frac1{4\pi}\sqrt\frac{E'(r_h)}{B^{(-1)}(r_h)},\\
  s&=\frac{C(r_h)^{d/2}}{4G_N}.
\end{split}\end{equation}

Chaos bound (\ref{ChaosBound}) is saturated in Einstein gravity. So
 \begin{equation}\label{LyapunovExp}
   \lambda_L=2\pi T.
 \end{equation}
We are going to derive the butterfly velocity $v_B$ in terms
of horizon quantities. Firstly we introduce the tortoise
coordinate
\begin{equation}
  dr_*=-\sqrt\frac {B(r)}{E(r)} dr
\end{equation}
to write the metric into
\begin{equation}
  ds^2=E(r)(-dt^2+dr_*^2)+C(r)d\textbf{x}^2,
\end{equation}
where the asymptotic boundary is located at $r_*=0$ and the horizon is
located at $r_*=-\infty$. Then we use the Kruskal coordinates
\begin{equation}
  uv=-e^{4\pi r_*/\beta},\quad u/v=-e^{-4\pi t/\beta},
\end{equation}
to further give
\begin{equation}
  ds^2= 2P(uv) dudv + Q(uv) d\textbf{x}^2,
\end{equation}
where $P(uv)=E(r)\frac2{uv}\left(\frac{\beta}{4\pi}\right)^2,\,
Q(uv)=C(r)$. The horizon is located at $uv=0$ and the
asymptotic boundary is located at $uv=-1$. One can find
\begin{equation}
  uv\sim(r-r_h)\theta + \cdots ,
\end{equation}
where $\theta>0$. Applying the expression of $v_B$ in
\cite{Roberts:2016wdl}, we can obtain
\begin{equation}\label{vBformula}
  v_B^2=\left(\frac{4\pi}{\beta}\right)^2
  \frac{P(0)}{2dQ'(0)}=\frac{E'(r_h)}{dC'(r_h)},
\end{equation}
where $P(0)=\frac{2 E'(r_h)}{\theta}\left(\frac{\beta}{4\pi}\right)^2$ and $ Q'(0)=\frac{C'(r_h)}{\theta}$ have been used.

We add the Maxwell term ${\cal S}_c$ (\ref{ElectroS}) into
the action to study the charge diffusion. Following the method in
\cite{Blake:2016wvh}, we firstly write down the DC conductivity $\sigma$
and the susceptibility $\chi^{-1}$
\begin{equation}
  \sigma=C(r_h)^{\frac d2-1},\quad \chi^{-1}=\int_0^{r_h} C(r)^{-\frac d2}\sqrt{B(r)E(r)} dr.
\end{equation}
Then the charge diffusion constant $D_c$ can be read from the Einstein relation
\begin{equation}\label{Dcformula}
  D_c=\frac{\sigma}{\chi}=C(r_h)^{\frac d2-1} \int_0^{r_h} C(r)^{-\frac d2}\sqrt{B(r)E(r)} dr.
\end{equation}
If above integration diverges, one can regularize it by
introducing a UV cutoff $\Lambda_\text{UV}$ into the integral as
$\int_{\Lambda_\text{UV}^{-1}}^{r_h}$.

\section{Numerical solutions for AdS-Schwarzschild black hole}\label{AppendixAdSNum}

Here we work on spatial dimension $d=2$. To study the
$AdS_4-AdS_4$ domain wall, we choose the potential as
\begin{equation}\label{AdSAdSPotential}
  V(\phi)=-6-\phi^2+\frac18\phi^4.
\end{equation}
We adopt the domain wall ansatz
\begin{equation}\label{AdSAdSAnsatz}
  ds^2={\cal F}(r)^{-1}dr^2+e^{2{\cal A}(r)}(-{\cal F}(r)dt^2+dx_1^2+dx_2^2),\quad \phi=\phi(r).
\end{equation}
The potential (\ref{AdSAdSPotential}) allows three $AdS_4$
fixed points. One of them has the larger radius of AdS and stays
in the UV ($r\to+\infty$), which is
\begin{equation}
  {\cal A}=r/L_\text{UV},\quad {\cal F}=1,\quad \phi=0
\end{equation}
with scalar modes
\begin{equation}
  \phi=\phi_-^\text{UV}e^{\Delta_-{\cal A}} + \phi_+^\text{UV}e^{\Delta_+{\cal A}},
\end{equation}
where $L_\text{UV}=1,\Delta_-=1,\Delta_+=2$. The other two
have the smaller radius of AdS and stay in the IR ($r\to-\infty$),
one of which is
\begin{equation}
  {\cal A}=r/L_\text{IR},\quad {\cal F}=1,\quad \phi=\phi_\text{IR}
\end{equation}
with scalar modes
\begin{equation}
  \phi=\phi_\text{IR}+\phi_-^\text{IR}e^{\delta_-{\cal A}} + \phi_+^\text{IR}e^{\delta_+{\cal A}},
\end{equation}
where
$L_\text{IR}=\frac{\sqrt3}{2},\phi_\text{IR}=2,\delta_\pm=\frac12(3\pm\sqrt{21})$.
The other IR fixed point is obtained by the reflection
$\phi\to-\phi$.

Firstly, we study the zero temperature flow. Given a slight
deviation from the IR fixed point, the irrelevant IR modes
$\phi_-^\text{IR}$ will be stimulated and intergraded to the UV
fixed point. We find a UV-IR relation
\begin{equation}\label{UVIRRelation}
  (\phi_-^\text{IR})^{1/\delta_-}=Z_\phi(\phi_-^\text{UV})^{1/\Delta_-},
\end{equation}
where the coefficient $Z_\phi=0.178$, which can be
understood as a renormalization of operator $\cal O_\phi$.

Secondly, we study the thermal flow. Notice that three sorts
of symmetries are contained in (\ref{AdSAdSAnsatz}): the first is
$r\to r+c$, which allows us to set the horizon at $r=0$; the
second is $r\to rc,t\to t/c,{\cal F}\to {\cal F}c^2$, which allows
us to set ${\cal F}'(0)=1$; the third is $t\to tc^2,x\to xc^2,
{\cal A}\to {\cal A}-\log c$, which allows us to set ${\cal
A}(0)=0$. Then, under the last boundary condition which sets
the value of $\phi(0)$, we can integrate the flow from the horizon
to the UV. Be cautious that ${\cal F}(+\infty)$ is no longer equal
to $1$ because of the gauge ${\cal F}'(0)=1$. One should recover
 ${\cal F}(+\infty)=1$ by using the second symmetry inversely.

Finally, according to (\ref{TFormula}), (\ref{vBformula}) and
(\ref{Dcformula}), we can numerically determine the
temperature $T$, butterfly velocity $v_B$ and diffusion ratios
$D_c\lambda_L/v_B^2$ by using
\begin{equation}
  T=\frac{e^{{\cal A}(0)} {\cal F}'(0)}{4 \pi },\quad v_B^2=\frac{{\cal F}'(0)}{4 {\cal A}'(0)},\quad
  \frac{D_c\lambda_L}{v_B^2} = 2 e^{{\cal A}(0)} {\cal A}'(0) \int_0^{\infty } e^{-{\cal A}(r)} \, dr.
\end{equation}

Let us employ the standard quantization and consider the
single trace deformation $W(\cal O_\phi)=\cal O_\phi$. Considering
the UV fixed point with relevant deformation, we can identify
$\kappa_s=\phi_-^\text{UV}$ in (\ref{vBdeform}), whose value can
be extracted by $\phi=\phi_-^\text{UV} e^{\Delta_-\cal A}+\cdots$ in
the UV. Considering the IR fixed point with irrelevant
deformation, we can identify $\kappa_s=\phi_-^\text{IR}$ in
(\ref{vBdeform}), whose value is given by (\ref{UVIRRelation}).

\begin{figure}
  \centering
  \includegraphics[width=200pt]{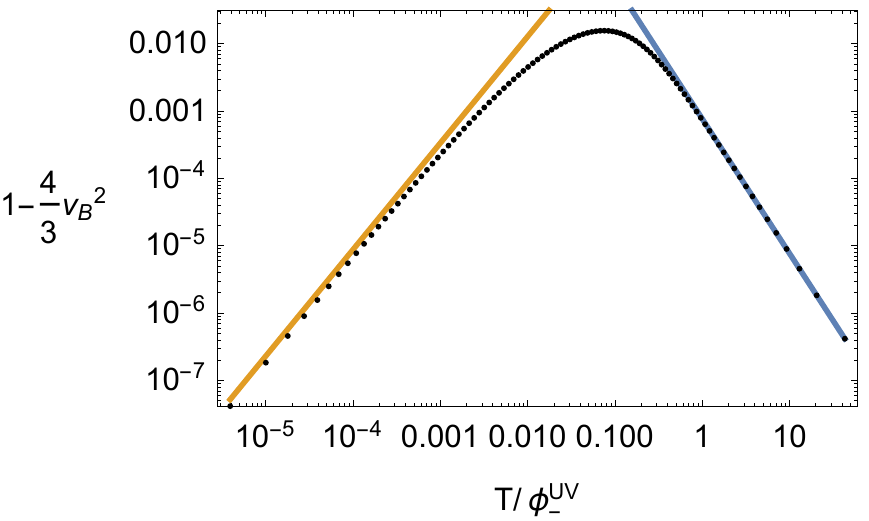}
  \includegraphics[width=200pt]{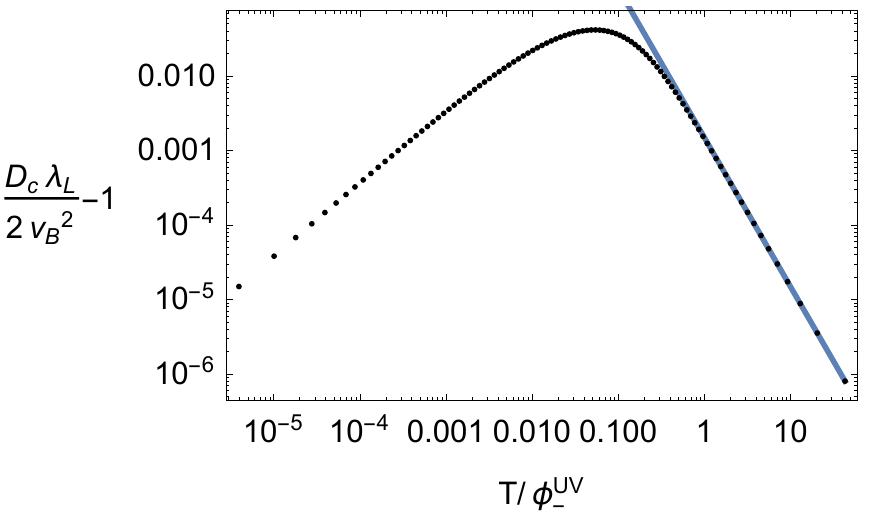}\\
  \caption{Quantities $1-\frac43v_B^2$ and $\frac{D_c\lambda_L}{2v_B^2}-1$ as functions of $T/\phi_-^\text{UV}$. Points are results of numerical calculation and lines are results of perturbation analysis.}\label{FigAdSAdSvB2Log}
\end{figure}

As $\Delta_-=1$, $T/\phi_-^\text{UV}$ is a dimensionless
parameter. In the left plot of Figure \ref{FigAdSAdSvB2Log}, we plot the numerical result of the quantity $1-\frac43v_B^2$ as a function of
$T/\phi_-^\text{UV}$. The analytical results based on (\ref{vBdeform}) for the UV fixed point and the IR fixed point are shown as well, where
the constant $\gamma$ of the UV (IR) fixed point is calculated by using $\Delta_-$ ($\delta_-$). The numerical result matches well with the analytical results
in both the $T\gg\phi_-^\text{UV}$ region and $T\ll\phi_-^\text{UV}$ region. The value of $v_B^2$ is also plotted in
the phase diagram in Figure \ref{FigQPT}. When
$\phi(0)\approx\phi_\text{IR}/2=1$, the effects of thermodynamic
and $\cal O_\phi$ deformation are commensurate. At this time,
$T/\phi_-^\text{UV}\approx0.1$ can be understood as the vague
boundary between the quantum critical region and the gapless low
temperature phase.

In the right plot of Figure \ref{FigAdSAdSvB2Log}, we plot the numerical result of the
quantity $\frac{D_c\lambda_L}{2v_B^2}-1$. The analytical result based on (\ref{Dcdeform}) for the UV fixed point is shown as well,
where the constant $\eta$ is calculated by using $\Delta_-$. Note
that the quantity $\frac{D_c\lambda_L}{2v_B^2}-1$ at low
$T/\phi_-^\text{UV}$ behaves as $T/\phi_-^\text{UV}$ rather than
$(T/\phi_-^\text{UV})^{-2\delta_-}$, which may result from the breakdown of
(\ref{AdSScalarMetricMode21}), because
the variation of the metric becomes divergent near asymptotic boundary of the IR
region under an irrelevant deformation.

\section{Numerical solutions for Lifshitz black hole}\label{AppendixLifNum}

For numerical calculation, we adopt the following ansatz
\begin{equation}\begin{split}
  ds^2&=\frac{L^2}{r^2}\left(-\frac{(1-r)U(r)}{r^{2z-2}}e^{-S(r)}dt^2 + \frac{dr^2}{(1-r)U(r)} + d\textbf{x}^2 \right),   \\
  {\cal B}&=L\sqrt\frac{2 (z-1)}{z} r^{-z} {\cal B}(r) (1-r)U(r)dt,\quad \phi=r^{\Delta_-}\phi(r),
\end{split}\end{equation}
and choose the parameters in the action (\ref{LifScalarAction}) as
\begin{equation}
  W=\frac{dz}{L^2},\quad V(\phi)= -\frac{z^2+z(d-1)+d^2}{L^2} + \frac{\Delta_-  (\Delta_- -(d+z))}{2 L^2}\phi^2.
\end{equation}
Then $L$ will not appear in the equation of motions. The
asymptotic boundary is located at $r\to0$ and the horizon is
located at $r=1$. The boundary conditions at $r=0$ are
$U(0)=1,S(0)=0,{\cal B}(0)=1,\phi(0)=\phi_-$, which ensure the
asymptotic Lifshitz solution (\ref{LifSpacetime}) with asymptotic
behavior (\ref{ScalarExpandr}). The boundary conditions at $r=1$
are regular conditions.

According to (\ref{TFormula}) and (\ref{vBformula}), temperature
$T$ and butterfly velocity $v_B$ in above ansatz are
separately given by
\begin{equation}
  T=\frac{ U(1)}{4 \pi }e^{-\frac{S(1)}{2}},\quad v_B^2= \frac{1}{2d} e^{-S(1)} U(1).
\end{equation}

Firstly, we build up a Lifshitz black hole without scalar
deformation by setting $\phi=0$. Then we calculate $\Phi(0)$ by
evaluating $v_B^2/T^{2-\frac2z}$, whose values, as a function of
$z$ in different spatial dimensions $d$, are shown in the left
plot of Figure \ref{FigLif}.

Secondly, we deform the Lifshitz black hole with scalar field. The
value of $\Delta_-$ should satisfy $0<\Delta_-<\frac{d+z}{2}$ to
make the deformation relevant. Then we can impose a small
perturbation with non-zero $\phi_-$ and study the variation of
$v_B$. The numerical results match (\ref{LifScalarvB}) when
$\phi_-\ll T^\frac{\Delta_-}{z}$, where the coefficient $\gamma$
as a function of $z$ for different $\Delta_-$ is shown in the
right plot of Figure \ref{FigLif}.

\section{Testing the relation between $v_\text{th}$ and $v_\text{E}$}\label{AppendixVthVE}

We consider a neutral black hole with flat horizon and scalar
hair. In \cite{Feng:2017wvc}, the authors propose a relation
between the thermodynamical volume density $v_\text{th}$ and the
Euclidean bounded volume density $v_\text{E}$ as
\begin{equation}\label{VthVE}
  v_\text{th}\doteq\sqrt\frac{g_1}{h_1} v_\text{E},
\end{equation}
where $g_1$ and $h_1$ appear in the near horizon expansion of the
black hole metric and the equality with a dot ``$\doteq$'' marks
their supposition. In a $\rho$ coordinate, the metric appears as
\begin{equation}\label{ansatzrho}
  ds^2=-h(\rho)dt^2+\frac{d\rho^2}{g(\rho)}+\frac{\rho^2}{L^2} d\textbf{x}^2
\end{equation}
with the horizon expansion
\begin{equation}
  h(\rho)\sim h_1(\rho-\rho_h)+h_2(\rho-\rho_h)^2+\cdots,\quad g(\rho)\sim g_1(\rho-\rho_h)+g_2(\rho-\rho_h)^2+\cdots
\end{equation}
and the asymptotic boundary expansion
\begin{equation}
  h(\rho)\sim \frac{\rho^2}{L^2}(1+\cdots-\frac{16 \pi G_N m_\text{BH} L^{d+2}}{d\rho^{d+1}}+\cdots),\quad g(\rho)\sim \frac{\rho^2}{L^2}(1+\cdots+\frac{\#}{\rho^{d+1}}+\cdots).
\end{equation}
The horizon is located at $\rho_h$ and the asymptotic boundary at $\rho\to\infty$. $m_\text{BH}$ is the mass density of the black hole. The Euclidean bounded volume density $v_\text{E}$ is
\begin{equation}
  v_E=\frac{1}{d+1}\rho_h^{d+1}L^{-d}.
\end{equation}
The thermodynamical volume density $v_\text{th}$ is determined by the general first law of black hole thermodynamics \cite{Kastor:2009wy,Cvetic:2010jb}
\begin{equation}\label{FirstLaw}
d m_\text{BH}=Tds+v_\text{th}dP,
\end{equation}
where the thermodynamical pressure of the black hole is
\begin{equation}
  P=\frac{d(d+1)}{16\pi G_N L^2}.
\end{equation}

For a neutral black hole with flat horizon and scalar hair, there are two Smarr relations  \cite{Feng:2017wvc} \footnote{Note that the coefficient before $v_\text{th}P$ is different from the one in \cite{Feng:2017wvc}. The reason is that the spatial component of our metric (\ref{ansatzrho}) is $\rho^2/L^2$, while the one in \cite{Feng:2017wvc} is $\rho^2$. Such difference change the definition of $v_\text{th}$ through the first law (\ref{FirstLaw}).}
\begin{eqnarray}\begin{split}
  m_\text{BH} &= \frac{d}{d+1}Ts,   \\
  m_\text{BH} &= \frac{d}{d-1}Ts-\frac4{d-1}v_\text{th}P,
\end{split}\end{eqnarray}
which give
\begin{equation}
  v_\text{th}P=\frac12m_\text{BH}=\frac{d}{2(d+1)}Ts.
\end{equation}
The hypothesis (\ref{VthVE}) relates the butterfly velocity $v_B$ to $v_\text{th}$ by
\begin{equation}\label{vBVth}
  v_B^2=\sqrt\frac{h_1}{g_1}\frac{Ts}{4v_\text{E} P}\doteq\frac{Ts}{4v_\text{th}P}=\frac{d+1}{2d},
\end{equation}
which leads to a constant $v_B$ and is in conflict with our
analytical result in (\ref{AdSScalarvB2r}) and numerical
result in Appendix \ref{AppendixAdSNum} where the scalar hair
is considered. Such contradiction results from the violation
of hypothesis (\ref{VthVE}) \footnote{Thank the authors in
\cite{Feng:2017wvc} for pointing out this.}.

Taking the coordinate transformation $\rho^2/L^2=C(\zeta)$, we
can change (\ref{AdSScalarSch}) into (\ref{ansatzrho}) and
check the hypothesis in (\ref{VthVE}). The result is
\begin{equation}
  \sqrt\frac{g_1}{h_1} \frac{v_\text{E}}{v_\text{th}}
  =1-\frac{d+1}{2 d} \lambda^2 (I_1(1;\vartheta)+2 I_2(1;\vartheta)),
\end{equation}
where (\ref{AdSScalarMetricMode10}), (\ref{AdSScalarMetricMode20})
and (\ref{AdSScalarMetricMode21}) have been used. In general, the last term
does not vanish. Especially, for
$0<\vartheta<\frac12$, it is non-positive. If further demanding one of the null-energy
conditions, $T^\rho{}_\rho-T^t{}_t\geq0$, we will have the inequality $\frac{g_1}{h_1}\geq1$ and finally obtain
\begin{equation}\label{VthVEIneq}
  v_\text{th}\geq\sqrt\frac{g_1}{h_1} v_\text{E}\geq v_\text{E}
\end{equation}
up to $O(\lambda^2)$, which is
the reverse isoperimetric inequality \cite{Feng:2017wvc}.

\end{appendix}


\begin{thebibliography}{79}

\bibitem{Forster:1975hyd}
  D.~Forster, ``Hydrodynamic fluctuations, broken symmetry, and correlation functions'', in Reading,
Mass., WA Benjamin, Inc.(Frontiers in Physics. Volume 47), 1975. 343 p., vol. 47, 1975.

\bibitem{Sekino:2008he}
  Y.~Sekino and L.~Susskind,
  ``Fast Scramblers,''  JHEP {\bf 0810}, 065 (2008)  
  [arXiv:0808.2096 [hep-th]].  

\bibitem{Hosur:2015ylk}
  P.~Hosur, X.~L.~Qi, D.~A.~Roberts and B.~Yoshida,
  ``Chaos in quantum channels,''  JHEP {\bf 1602}, 004 (2016)  
  [arXiv:1511.04021 [hep-th]].  

\bibitem{Shenker:2013pqa}
  S.~H.~Shenker and D.~Stanford,
  ``Black holes and the butterfly effect,''  JHEP {\bf 1403}, 067 (2014)  
  [arXiv:1306.0622 [hep-th]].  

\bibitem{Sircar:2016old}
  N.~Sircar, J.~Sonnenschein and W.~Tangarife,
  ``Extending the scope of holographic mutual information and chaotic behavior,''
  JHEP {\bf 1605}, 091 (2016)
  [arXiv:1602.07307 [hep-th]].

\bibitem{Cai:2017ihd}
  R.~G.~Cai, X.~X.~Zeng and H.~Q.~Zhang,
  ``Influence of inhomogeneities on holographic mutual information and butterfly effect,''
  arXiv:1704.03989 [hep-th].

\bibitem{Maldacena:2015waa}
  J.~Maldacena, S.~H.~Shenker and D.~Stanford,
  ``A bound on chaos,''  JHEP {\bf 1608}, 106 (2016)  
  [arXiv:1503.01409 [hep-th]].  

\bibitem{Shenker:2013yza}
  S.~H.~Shenker and D.~Stanford,
  ``Multiple Shocks,''  JHEP {\bf 1412}, 046 (2014)  
  [arXiv:1312.3296 [hep-th]].  

\bibitem{Roberts:2014isa}
  D.~A.~Roberts, D.~Stanford and L.~Susskind,
  ``Localized shocks,''  JHEP {\bf 1503}, 051 (2015)  
  [arXiv:1409.8180 [hep-th]].  

\bibitem{Kitaev:2014hch}
  A.~Kitaev, ``Hidden correlations in the hawking radiation and thermal noise,'' (2014), talk given at the Fundamental Physics Prize Symposium, Nov. 10, 2014.

\bibitem{Roberts:2014ifa}
  D.~A.~Roberts and D.~Stanford,
  ``Two-dimensional conformal field theory and the butterfly effect,''  Phys.\ Rev.\ Lett.\  {\bf 115}, no. 13, 131603 (2015)  
  [arXiv:1412.5123 [hep-th]].  

\bibitem{Stanford:2015owe}
  D.~Stanford,
  ``Many-body chaos at weak coupling,''  JHEP {\bf 1610}, 009 (2016)  
  [arXiv:1512.07687 [hep-th]].  

\bibitem{Polchinski:2016xgd}
  J.~Polchinski and V.~Rosenhaus,
  ``The Spectrum in the Sachdev-Ye-Kitaev Model,''  JHEP {\bf 1604}, 001 (2016)  
  [arXiv:1601.06768 [hep-th]].  

\bibitem{Roberts:2016wdl}
  D.~A.~Roberts and B.~Swingle,
  ``Lieb-Robinson Bound and the Butterfly Effect in Quantum Field Theories,''  Phys.\ Rev.\ Lett.\  {\bf 117}, no. 9, 091602 (2016)  
  [arXiv:1603.09298 [hep-th]].  

\bibitem{Maldacena:2016hyu}
  J.~Maldacena and D.~Stanford,
  ``Remarks on the Sachdev-Ye-Kitaev model,''  Phys.\ Rev.\ D {\bf 94}, no. 10, 106002 (2016)  
  [arXiv:1604.07818 [hep-th]].  

\bibitem{Kovtun:2004de}
  P.~Kovtun, D.~T.~Son and A.~O.~Starinets,
  ``Viscosity in strongly interacting quantum field theories from black hole physics,''
  Phys.\ Rev.\ Lett.\  {\bf 94}, 111601 (2005)
  [hep-th/0405231].

\bibitem{Swingle:2016var}
  B.~Swingle, G.~Bentsen, M.~Schleier-Smith and P.~Hayden,
  ``Measuring the scrambling of quantum information,''  Phys.\ Rev.\ A {\bf 94}, no. 4, 040302 (2016)  
  [arXiv:1602.06271 [quant-ph]].  

\bibitem{Hartnoll:2014lpa}
  S.~A.~Hartnoll,
  ``Theory of universal incoherent metallic transport,''  Nature Phys.\  {\bf 11}, 54 (2015)  
  [arXiv:1405.3651 [cond-mat.str-el]].  

\bibitem{Blake:2016wvh}
  M.~Blake,
  ``Universal Charge Diffusion and the Butterfly Effect in Holographic Theories,''  Phys.\ Rev.\ Lett.\  {\bf 117}, no. 9, 091601 (2016)  
  [arXiv:1603.08510 [hep-th]].  

\bibitem{Blake:2016sud}
  M.~Blake,
  ``Universal Diffusion in Incoherent Black Holes,''  Phys.\ Rev.\ D {\bf 94}, no. 8, 086014 (2016)  
  [arXiv:1604.01754 [hep-th]].  

\bibitem{Lucas:2016yfl}
  A.~Lucas and J.~Steinberg,
  ``Charge diffusion and the butterfly effect in striped holographic matter,''
  JHEP {\bf 1610}, 143 (2016)
  [arXiv:1608.03286 [hep-th]].

\bibitem{Patel:2016wdy}
  A.~A.~Patel and S.~Sachdev,
  ``Quantum chaos on a critical Fermi surface,''  Proc.\ Nat.\ Acad.\ Sci.\  {\bf 114}, 1844 (2017)  
  [arXiv:1611.00003 [cond-mat.str-el]].  

\bibitem{Blake:2016jnn}
  M.~Blake and A.~Donos,
  ``Diffusion and Chaos from near AdS$_2$ horizons,''  JHEP {\bf 1702}, 013 (2017)  
  [arXiv:1611.09380 [hep-th]].  

\bibitem{Baggioli:2016pia}
  M.~Baggioli, B.~Gout\'eraux, E.~Kiritsis and W.~J.~Li,
  ``Higher derivative corrections to incoherent metallic transport in holography,''  JHEP {\bf 1703}, 170 (2017)
  [arXiv:1612.05500 [hep-th]].  

\bibitem{Kim:2017dgz}
  K.~Y.~Kim and C.~Niu,
  ``Diffusion and Butterfly Velocity at Finite Density,''
  arXiv:1704.00947 [hep-th].

\bibitem{Baggioli:2017ojd}
  M.~Baggioli and W.~J.~Li,
  ``Diffusivities bounds and chaos in holographic Horndeski theories,''
  arXiv:1705.01766 [hep-th].

\bibitem{Blake:2017qgd}
  M.~Blake, R.~A.~Davison and S.~Sachdev,
  ``Thermal diffusivity and chaos in metals without quasiparticles,''  arXiv:1705.07896 [hep-th].  

\bibitem{Hartman:2017hhp}
  T.~Hartman, S.~A.~Hartnoll and R.~Mahajan,
  ``An upper bound on transport,''
  arXiv:1706.00019 [hep-th].

\bibitem{Gu:2016oyy}
  Y.~Gu, X.~L.~Qi and D.~Stanford,
  ``Local criticality, diffusion and chaos in generalized Sachdev-Ye-Kitaev models,''  JHEP {\bf 1705}, 125 (2017)  
  [arXiv:1609.07832 [hep-th]].  

\bibitem{Davison:2016ngz}
  R.~A.~Davison, W.~Fu, A.~Georges, Y.~Gu, K.~Jensen and S.~Sachdev,
  ``Thermoelectric transport in disordered metals without quasiparticles: The Sachdev-Ye-Kitaev models and holography,''  Phys.\ Rev.\ B {\bf 95}, no. 15, 155131 (2017)  
  [arXiv:1612.00849 [cond-mat.str-el]].  

\bibitem{Gu:2017ohj}
  Y.~Gu, A.~Lucas and X.~L.~Qi,
  ``Energy diffusion and the butterfly effect in inhomogeneous Sachdev-Ye-Kitaev chains,''  SciPost Phys.\  {\bf 2}, 018 (2017)  
  [arXiv:1702.08462 [hep-th]].  

\bibitem{Shen:2016htm}
  H.~Shen, P.~Zhang, R.~Fan and H.~Zhai,
  ``Out-of-Time-Order Correlation at a Quantum Phase Transition,''  arXiv:1608.02438 [cond-mat.str-el].  

\bibitem{Bohrdt:2016vhv}
  A.~Bohrdt, C.~B.~Mendl, M.~Endres and M.~Knap,
  ``Scrambling and thermalization in a diffusive quantum many-body system,''  New J.\ Phys.\  {\bf 19}, no. 6, 063001 (2017)  
  [arXiv:1612.02434 [cond-mat.quant-gas]].  

\bibitem{Chowdhury:2017jzb}
  D.~Chowdhury and B.~Swingle,
  ``Onset of many-body chaos in the $O(N)$ model,''  arXiv:1703.02545 [cond-mat.str-el].  

\bibitem{Sachdev:1999QPT}
  S. Sachdev, ``Quantum Phase Transitions,'' 2nd Edition, Cambridge University Press (2011).

\bibitem{Hartnoll:2009sz}
  S.~A.~Hartnoll,
  ``Lectures on holographic methods for condensed matter physics,''  Class.\ Quant.\ Grav.\  {\bf 26}, 224002 (2009)  
  [arXiv:0903.3246 [hep-th]].  

\bibitem{Sachdev:2010ch}
  S.~Sachdev,
  ``Condensed Matter and AdS/CFT,''  Lect.\ Notes Phys.\  {\bf 828}, 273 (2011)  
  [arXiv:1002.2947 [hep-th]].  

\bibitem{Hartnoll:2016apf}
  S.~A.~Hartnoll, A.~Lucas and S.~Sachdev,
  ``Holographic quantum matter,''
  arXiv:1612.07324 [hep-th].  

\bibitem{Lucas:2017dqa}
  A.~Lucas, T.~Sierens and W.~Witczak-Krempa,
  ``Quantum critical response: from conformal perturbation theory to holography,''
  arXiv:1704.05461 [hep-th].

\bibitem{Zaanen:2004sth}
  J.~Zaanen,
  ``Superconductivity: Why the temperature is high,''
  Nature 430 (2004) 512.

\bibitem{Dong:2012se}
  X.~Dong, S.~Harrison, S.~Kachru, G.~Torroba and H.~Wang,
  ``Aspects of holography for theories with hyperscaling violation,''
  JHEP {\bf 1206}, 041 (2012)
  [arXiv:1201.1905 [hep-th]].

\bibitem{Gouteraux:2011ce}
  B.~Gout\'eraux and E.~Kiritsis,
  ``Generalized Holographic Quantum Criticality at Finite Density,''
  JHEP {\bf 1112}, 036 (2011)
  [arXiv:1107.2116 [hep-th]].

\bibitem{Gouteraux:2011qh}
  B.~Gout\'eraux, J.~Smolic, M.~Smolic, K.~Skenderis and M.~Taylor,
  ``Holography for Einstein-Maxwell-dilaton theories from generalized dimensional reduction,''
  JHEP {\bf 1201}, 089 (2012)
  [arXiv:1110.2320 [hep-th]].

\bibitem{McGreevy:2009xe}
  J.~McGreevy,
  ``Holographic duality with a view toward many-body physics,''  Adv.\ High Energy Phys.\  {\bf 2010}, 723105 (2010)  
  [arXiv:0909.0518 [hep-th]].  

\bibitem{Ling:2016ibq}
  Y.~Ling, P.~Liu and J.~P.~Wu,
  ``Holographic Butterfly Effect at Quantum Critical Points,''
  arXiv:1610.02669 [hep-th].

\bibitem{Feng:2017wvc}
  X.~H.~Feng and H.~Lu,
  ``Butterfly Velocity Bound and Reverse Isoperimetric Inequality,''  Phys.\ Rev.\ D {\bf 95}, no. 6, 066001 (2017)  
  [arXiv:1701.05204 [hep-th]].  

\bibitem{Witten:2001ua}
  E.~Witten,
  ``Multitrace operators, boundary conditions, and AdS / CFT correspondence,''  hep-th/0112258.  

\bibitem{Gubser:2002zh}
  S.~S.~Gubser and I.~Mitra,
  ``Double trace operators and one loop vacuum energy in AdS / CFT,''  Phys.\ Rev.\ D {\bf 67}, 064018 (2003)  
  [hep-th/0210093].  

\bibitem{Faulkner:2010gj}
  T.~Faulkner, G.~T.~Horowitz and M.~M.~Roberts,
  ``Holographic quantum criticality from multi-trace deformations,''  JHEP {\bf 1104}, 051 (2011)  
  [arXiv:1008.1581 [hep-th]].  

\bibitem{Faulkner:2010fh}
  T.~Faulkner, G.~T.~Horowitz and M.~M.~Roberts,
  ``New stability results for Einstein scalar gravity,''  Class.\ Quant.\ Grav.\  {\bf 27}, 205007 (2010)  
  [arXiv:1006.2387 [hep-th]].  

\bibitem{Ling:2016wuy}
  Y.~Ling, P.~Liu and J.~P.~Wu,
  ``Note on the butterfly effect in holographic superconductor models,''
  arXiv:1610.07146 [hep-th].

\bibitem{Gubser:2009cg}
  S.~S.~Gubser and A.~Nellore,
  ``Ground states of holographic superconductors,''  Phys.\ Rev.\ D {\bf 80}, 105007 (2009)  
  [arXiv:0908.1972 [hep-th]].  

\bibitem{Caldarelli:2016nni}
  M.~M.~Caldarelli, A.~Christodoulou, I.~Papadimitriou and K.~Skenderis,
  ``Phases of planar AdS black holes with axionic charge,''  JHEP {\bf 1704}, 001 (2017)  
  [arXiv:1612.07214 [hep-th]].  

\bibitem{Kachru:2008yh}
  S.~Kachru, X.~Liu and M.~Mulligan,
  ``Gravity duals of Lifshitz-like fixed points,''
  Phys.\ Rev.\ D {\bf 78}, 106005 (2008)
  [arXiv:0808.1725 [hep-th]].

\bibitem{Taylor:2008tg}
  M.~Taylor,
  ``Non-relativistic holography,''
  arXiv:0812.0530 [hep-th].

\bibitem{Taylor:2015glc}
  M.~Taylor,
  ``Lifshitz holography,''  Class.\ Quant.\ Grav.\  {\bf 33}, no. 3, 033001 (2016)  
  [arXiv:1512.03554 [hep-th]].  

\bibitem{Sachdev:2008ba}
  S.~Sachdev and M.~Mueller,
  ``Quantum criticality and black holes,''
  J.\ Phys.\ Condens.\ Matter {\bf 21}, 164216 (2009)
  [arXiv:0810.3005 [cond-mat.str-el]].

\bibitem{Charmousis:2010zz}
  C.~Charmousis, B.~Gout\'eraux, B.~S.~Kim, E.~Kiritsis and R.~Meyer,
  ``Effective Holographic Theories for low-temperature condensed matter systems,''
  JHEP {\bf 1011}, 151 (2010)
  [arXiv:1005.4690 [hep-th]].

\bibitem{Gouteraux:2012yr}
  B.~Gout\'eraux and E.~Kiritsis,
  ``Quantum critical lines in holographic phases with (un)broken symmetry,''
  JHEP {\bf 1304}, 053 (2013)
  [arXiv:1212.2625 [hep-th]].

\bibitem{Kiritsis:2015oxa}
  E.~Kiritsis and J.~Ren,
  ``On Holographic Insulators and Supersolids,''
  JHEP {\bf 1509}, 168 (2015)
  [arXiv:1503.03481 [hep-th]].

\bibitem{Liu:2013una}
  H.~Liu and M.~Mezei,
  ``Probing renormalization group flows using entanglement entropy,''  JHEP {\bf 1401}, 098 (2014)  
  [arXiv:1309.6935 [hep-th]].  

\bibitem{Gursoy:2008za}
  U.~Gursoy, E.~Kiritsis, L.~Mazzanti and F.~Nitti,
  ``Holography and Thermodynamics of 5D Dilaton-gravity,''
  JHEP {\bf 0905}, 033 (2009)
  [arXiv:0812.0792 [hep-th]].

\bibitem{Ling:2016ien}
  Y.~Ling, Z.~Y.~Xian and Z.~Zhou,
  ``Holographic Shear Viscosity in Hyperscaling Violating Theories without Translational Invariance,''  JHEP {\bf 1611}, 007 (2016)  
  [arXiv:1605.03879 [hep-th]].  

\bibitem{Skenderis:2002wp}
  K.~Skenderis,
  ``Lecture notes on holographic renormalization,''
  Class.\ Quant.\ Grav.\  {\bf 19}, 5849 (2002)
  [hep-th/0209067].

\bibitem{Anabalon:2015xvl}
  A.~Anabalon, D.~Astefanesei, D.~Choque and C.~Martinez,
  ``Trace Anomaly and Counterterms in Designer Gravity,''  JHEP {\bf 1603}, 117 (2016)
  [arXiv:1511.08759 [hep-th]].  

\bibitem{Astefanesei:2008wz}
  D.~Astefanesei, N.~Banerjee and S.~Dutta,
  ``(Un)attractor black holes in higher derivative AdS gravity,''  JHEP {\bf 0811}, 070 (2008)  
  [arXiv:0806.1334 [hep-th]].  

\bibitem{Kaplan:2009kr}
  D.~B.~Kaplan, J.~W.~Lee, D.~T.~Son and M.~A.~Stephanov,
  ``Conformality Lost,''  Phys.\ Rev.\ D {\bf 80}, 125005 (2009)  
  [arXiv:0905.4752 [hep-th]].  

\bibitem{Iqbal:2010eh}
  N.~Iqbal, H.~Liu, M.~Mezei and Q.~Si,
  ``Quantum phase transitions in holographic models of magnetism and superconductors,''  Phys.\ Rev.\ D {\bf 82}, 045002 (2010)  
  [arXiv:1003.0010 [hep-th]].  

\bibitem{Jensen:2010ga}
  K.~Jensen, A.~Karch, D.~T.~Son and E.~G.~Thompson,
  ``Holographic Berezinskii-Kosterlitz-Thouless Transitions,''
  Phys.\ Rev.\ Lett.\  {\bf 105}, 041601 (2010)
  [arXiv:1002.3159 [hep-th]].

\bibitem{Iqbal:2011ae}
  N.~Iqbal, H.~Liu and M.~Mezei,
  ``Lectures on holographic non-Fermi liquids and quantum phase transitions,''  arXiv:1110.3814 [hep-th].  

\bibitem{Evans:2010hi}
  N.~Evans, A.~Gebauer, K.~Y.~Kim and M.~Magou,
  ``Phase diagram of the D3/D5 system in a magnetic field and a BKT transition,''  Phys.\ Lett.\ B {\bf 698}, 91 (2011)  
  [arXiv:1003.2694 [hep-th]].  

\bibitem{Iqbal:2011aj}
  N.~Iqbal, H.~Liu and M.~Mezei,
  ``Quantum phase transitions in semilocal quantum liquids,''  Phys.\ Rev.\ D {\bf 91}, no. 2, 025024 (2015)  
  [arXiv:1108.0425 [hep-th]].  

\bibitem{Jensen:2010vx}
  K.~Jensen,
  ``More Holographic Berezinskii-Kosterlitz-Thouless Transitions,''  Phys.\ Rev.\ D {\bf 82}, 046005 (2010)  
  [arXiv:1006.3066 [hep-th]].  

\bibitem{Freedman:1999gp}
  D.~Z.~Freedman, S.~S.~Gubser, K.~Pilch and N.~P.~Warner,
  ``Renormalization group flows from holography supersymmetry and a c theorem,''  Adv.\ Theor.\ Math.\ Phys.\  {\bf 3}, 363 (1999)  [hep-th/9904017].  

\bibitem{Kastor:2009wy}
  D.~Kastor, S.~Ray and J.~Traschen,
  ``Enthalpy and the Mechanics of AdS Black Holes,''  Class.\ Quant.\ Grav.\  {\bf 26}, 195011 (2009)
  [arXiv:0904.2765 [hep-th]].  

\bibitem{Cvetic:2010jb}
  M.~Cvetic, G.~W.~Gibbons, D.~Kubiznak and C.~N.~Pope,
  ``Black Hole Enthalpy and an Entropy Inequality for the Thermodynamic Volume,''  Phys.\ Rev.\ D {\bf 84}, 024037 (2011)  
  [arXiv:1012.2888 [hep-th]].  

\bibitem{Qaemmaqami:2017jxz}
  M.~M.~Qaemmaqami,
  ``On the Butterfly Effect in 3D Gravity,''
  arXiv:1707.00509 [hep-th].

\bibitem{Alishahiha:2016cjk}
  M.~Alishahiha, A.~Davody, A.~Naseh and S.~F.~Taghavi,
  ``On Butterfly effect in Higher Derivative Gravities,''  JHEP {\bf 1611}, 032 (2016)  
  [arXiv:1610.02890 [hep-th]].  

\bibitem{Qaemmaqami:2017bdn}
  M.~M.~Qaemmaqami,
  ``Criticality in Third Order Lovelock Gravity and Butterfly effect,''
  arXiv:1705.05235 [hep-th].



\end{thebibliography}
\end{document}